\documentclass[preprint]{aastex}

\begin{document}
\title{The ACS LCID project. V. The Star Formation History of the Dwarf Galaxy \objectname[]{LGS-3}: 
Clues for Cosmic Reionization and Feedback\altaffilmark{1}}

\author{Sebastian L. Hidalgo\altaffilmark{2,3},
Antonio Aparicio\altaffilmark{3,2},
Evan Skillman\altaffilmark{4},
Matteo Monelli\altaffilmark{2,3},
Carme Gallart\altaffilmark{2,3},
Andrew Cole\altaffilmark{5},
Andrew Dolphin\altaffilmark{6},
Daniel Weisz\altaffilmark{4},
Edouard Bernard\altaffilmark{7},
Santi Cassisi\altaffilmark{8},
Lucio Mayer\altaffilmark{9,10},
Peter Stetson\altaffilmark{11},
Eline Tolstoy\altaffilmark{12}, and
Henry Ferguson\altaffilmark{13}}

\altaffiltext{1}{Based on observations made with the NASA/ESA Hubble Space Telescope, 
obtained at the Space Telescope Science Institute, which is operated by the 
Association of Universities for Research in Astronomy, Inc., under NASA contract NAS 
5-26555. These observations are associated with program \#10505}
\altaffiltext{2}{Instituto de Astrof\'\i sica de Canarias. V\'\i a L\'actea s/n.
E38200 - La Laguna, Tenerife, Canary Islands, Spain;
shidalgo@iac.es, aparicio@iac.es, monelli@iac.es, carme@iac.es}
\altaffiltext{3}{Department of Astrophysics, University of La Laguna. V\'\i a L\'actea s/n.
E38200 - La Laguna, Tenerife, Canary Islands, Spain}
\altaffiltext{4}{Astronomy Department, University of Minnesota, Minneapolis, MN, USA;
skillman@astro.umn.edu, dweisz@astro.umn.edu}
\altaffiltext{5}{School of Mathematics \& Physics, University of Tasmania,
Hobart, Tasmania, Australia; andrew.cole@utas.edu.au}
\altaffiltext{6}{Raytheon; 1151 E. Hermans Rd., Tucson, AZ 85706, USA;
adolphin@raytheon.com}
\altaffiltext{7}{Institute for Astronomy, University of Edinburgh, Royal
Observatory, Blackford Hill, Edinburgh EH9 3HJ, UK; ejb@roe.ac.uk}
\altaffiltext{8}{INAF-Osservatorio Astronomico di Collurania,
Teramo, Italy; cassisi@oa-teramo.inaf.it}
\altaffiltext{9}{Institut f\"ur Theoretische Physik, University of Zurich,
Z\"urich, Switzerland; lucio@physik.unizh.ch}
\altaffiltext{10}{Department of Physics, Institut f\"ur Astronomie,
ETH Z\"urich, Z\"urich, Switzerland; lucio@phys.ethz.ch}
\altaffiltext{11}{Dominion Astrophysical Observatory, Herzberg Institute of
Astrophysics, National Research Council, 5071 West Saanish Road, Victoria,
British Columbia V9E 2E7, Canada; peter.stetson@nrc-cnrc.gc.ca}
\altaffiltext{12}{Kapteyn Astronomical Institute, University of Groningen,
    Groningen, Netherlands; etolstoy@astro.rug.nl}
\altaffiltext{13}{Space Telescope Science Institute, 3700 San Martin Drive, Baltimore, MD 21218, USA; 
ferguson@stsci.edu}

\begin{abstract}

We present an analysis of the star formation history (SFH) of the transition-type 
(dIrr/dSph) Local Group galaxy \objectname[]{LGS-3} (Pisces) based on deep photometry obtained with the
{\it Advanced Camera for Surveys} onboard the {\it Hubble Space Telescope}. 
Our observations 
reach the oldest main sequence turn-offs at high S/N, allowing a time 
resolution at the oldest ages of $\sigma\sim 1.1$ Gyr. 
Our analysis, based on three different SFH codes, shows that the 
SFH of \objectname[]{LGS-3} is dominated by a main episode $\sim 11.7$ Gyr ago with a duration of 
$\sim$ 1.4 Gyr. 
Subsequently, \objectname[]{LGS-3} continued forming stars until the present, 
although at a much lower rate. 
Roughly 90\% of the stars in \objectname[]{LGS-3} were formed in the initial episode
of star formation.
Extensive tests of self-consistency, uniqueness, and stability of the solution 
have been performed together with the IAC-star/IAC-pop/MinnIAC codes and these
results are found to be independent of the photometric reduction package, 
the stellar evolution library, and the SFH recovery method.
There is little evidence of chemical enrichment during the initial episode of 
star formation, after which the metallicity increased more steeply reaching a 
present day value of $Z \sim 0.0025$. 
This suggests a scenario in which \objectname[]{LGS-3} first formed stars mainly from 
infalling fresh gas, and after about $9$ Gyr ago, from a larger fraction of 
recycled gas.
The lack of early chemical enrichment is in contrast to that observed 
in the isolated dSph galaxies of comparable luminosity, implying that
the dSphs were more massive and subjected to more tidal stripping.

We compare the SFH of \objectname[]{LGS-3} with expectations from cosmological models.  
Most or all the star formation was produced in \objectname[]{LGS-3} after the reionization epoch, 
assumed to be completed at $z\sim6$ or $\sim 12.7$ Gyr ago. The total mass of the 
galaxy is estimated to be between 2 and $4\times 10^8$ M$_\odot$,
corresponding to circular velocities between 28 km\ s$^{-1}$ to 36 km\ s$^{-1}$. 
These values are close to but somewhat above the limit of 30 km\ s$^{-1}$ below 
which the UV background is expected to prevent any star formation after 
reionization. 
Feedback from SNe associated with the initial episode of star formation  
(mechanical luminosity from SNe $L_w=5.3\times 10^{38}$ erg s$^{-1}$) is probably
inadequate to completely blow away the gas. However, the combined effects of 
SN feedback and UV background heating might be expected to completely halt 
star formation at the reionization epoch for the low mass of \objectname[]{LGS-3}; this suggests 
that self-shielding is important to the early evolution of galaxies in 
this mass range.

\end{abstract}


\section{Introduction}\label{secint}

The present paper is part of the Local Cosmology from Isolated Dwarfs 
(LCID)\footnote{Local Cosmology from Isolated Dwarfs: http://www.iac.es/project/LCID/} project. 
The Hubble Space Telescope (HST) has been used to obtain deep photometry of six isolated dwarf galaxies in the Local Group: IC 1613, Leo A, Cetus, Tucana, \objectname[]{LGS-3} and Phoenix. Color-magnitude diagrams (CMD) 
reaching the oldest turn-off points have been derived. The Advanced Camera for Surveys \citep[ACS,][]{ford98} has been 
used to observe the first five, while Phoenix was observed with the Wide Field and 
Planetary Camera-2 (WFPC2). The main goal of the LCID project is to derive the 
star formation history (SFH), age-metallicity relation (AMR), and stellar population 
gradients of this sample of galaxies. Our objective is to study their evolution 
at early epochs and to probe effects of cosmological processes, such as the cosmic 
UV background subsequent to the onset of star formation in the universe or 
physical processes such as the gas removal by supernovae (SNe) feedback. 
Our sample consists of field dwarfs which were chosen in an effort to study systems 
as free as possible from environmental effects due to strong interactions with 
a host, massive galaxy. The SFH is a powerful tool to derive fundamental 
properties of these systems and their evolution, but to study the earliest 
epochs of star formation, deep CMDs, reaching the oldest main sequence turn-offs,
are required.

Here we present our analysis of \objectname[]{LGS-3}.
\objectname[]{LGS-3} is a dIrr/dSph low-luminosity galaxy discovered by \citet{kar1976}. 
From its location, it appears to be a member of the Andromeda subgroup of 
the Local Group \citep{ber2000}. The distance to the galaxy has been estimated 
by several authors. \citet{lee1995}, \citet{tik_mak1996}, \citet{mou1997}, 
and \citet*{apa_gal_ber1997} used the tip of the red giant branch (TRGB) to 
estimate distances ranging from 0.6 to 0.96 Mpc. This measurement is
inherently challenging because the TRGB region of \objectname[]{LGS-3} is very sparsely populated. 
\citet{mil_etal2001} also used the TRGB but complemented with observations of the 
horizontal branch (HB), the red clump (RC), and the overall match of the CMD 
to model CMDs to estimate a distance of $0.62\pm 0.02$ Mpc. Finally, 
from a study of the RR-Lyr stars in LGS-3, \citet{ber2009} derived a distance 
of $0.65\pm 0.05$ Mpc ($(m-M)_0 =24.07\pm 0.15$  mag). We will assume this value in this paper. 

The stellar content of \objectname[]{LGS-3} has been studied by \citet{lee1995}, \citet{mou1997}, 
\citet{apa_gal_ber1997}, and \citet{mil_etal2001}. \citet{mou1997} discovered the first Cepheid 
variable candidates in this galaxy. These works show that \objectname[]{LGS-3} 
contains stars of all ages; it has been forming stars since at least 12 Gyr ago 
and shows a low recent star formation rate (SFR) \citep{apa_gal_ber1997}. 
The SFH shows a dependency with radius \citep{mil_etal2001}.

No \ion{H}{2} regions have been found in the galaxy \citep{hod_mil1995}, which 
is consistent with the low recent star formation activity. The metallicity of 
the stars has been estimated from the color of the red giant branch (RGB). 
In this way, \citet{lee1995} obtained a mean metallicity of 
$\rm [Fe/H]=-2.10\pm 0.22$, while \citet{apa_gal_ber1997} derived a metallicity 
range $\rm -1.45\leq [M/H]\leq -1.0$. 
\citet{mil_etal2001} derived $\rm [Fe/H]=-1.5\pm 0.3$ for the stars older than 8 Gyr, 
and $\rm [Fe/H]\sim -1$ for the most recent generation of stars, 

The HI content of \objectname[]{LGS-3} was first studied with single dish spectra by \citet{thu_mar1979} 
and, more recently, with interferometry by \citet{you_lo1997}. 
Using the results by the latter authors and the 
adopted distance of 0.65 Mpc, the results from \citet{you_lo1997}
correspond to a total HI mass of $M_{HI} =2.7\times 10^5$ M$_\odot$, which 
corresponds to a total atomic gas mass of $3.8\times 10^5$ M$_\odot$ when corrected 
for helium. They also measured a gas velocity dispersion of 8 km\ s$^{-1}$ 
and negligible rotation at a galactocentric radius of 470 pc. The corresponding 
virial mass estimate is $M_{vir}=2.1\times 10^7$ M$_\odot$. 

In this paper, we present the SFH of \objectname[]{LGS-3} obtained from observations with the ACS 
on the HST. The photometry reaches the oldest main sequence turn-offs of the galaxy, 
allowing us to obtain an accurate SFH even for the oldest stellar populations. 
The structure of the paper is as follows: in \S\ref{secred} the observations and 
data reduction are discussed, the CMD is presented, and the completeness tests are 
described.  In \S\ref{secsfhmethod} we describe many of the details of the 
IAC-pop/MinnIAC methodology and its application to the LCID program 
observations.  The derived SFH of \objectname[]{LGS-3} is presented in \S\ref{secsfhresult} 
and is compared with those of other LCID galaxies in \S\ref{seclcid}. 
The consequences of SNe driven winds and cosmic reionization on the SFH of \objectname[]{LGS-3} 
are discussed in \S\ref{seccosmo}. The main conclusions of the work are summarized 
in \S\ref{seccon}. 
Finally, in appendix \ref{secimf}, the uniqueness, 
self-consistency, stability, and statistical significance of the solution 
are discussed. 

\section{Observations and Data Reduction}\label{secred}

\begin{figure}
\centering
\includegraphics[width=14cm,angle=0]{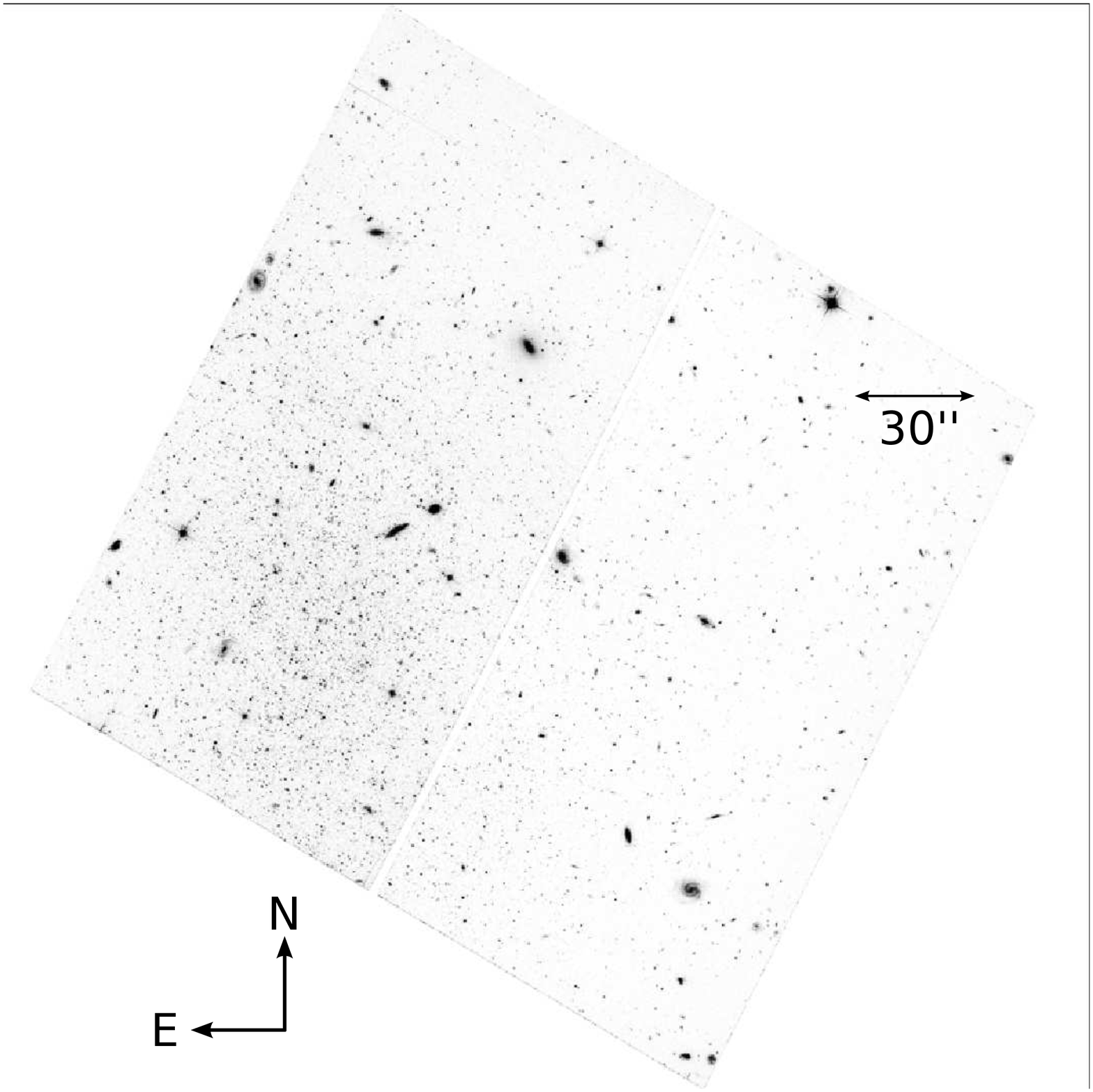}
\protect\caption[ ]{The LGS-3 observed field. Orientation and scale are marked.
\label{f1}}
\end{figure}

The ACS observations of \objectname[]{LGS-3} were obtained between September 12 and 17, 2005. 
The F475W and F814W bands were selected as the most efficient combination to trace
age differences at old ages, since they provide the
smallest relative error in age and metallicity in the main-sequence and sub-giant 
regions.
Total integration times were 15072 s in F475W and
13824 s in F814W.  The observations were organized in 6 visits of 2 orbits
each, and each orbit was split into one F475W and one F814W exposure. In the
first orbit of each visit, exposure times were 1250 and 1147 s respectively while in
the second orbit they were 1262 and 1157 s
respectively. Dithers of a few pixels between exposures were introduced to
minimize the impact of pixel-to-pixel sensitivity variations (``hot pixels'') 
in the CCDs. The observed field of \objectname[]{LGS-3} is
shown in Figure \ref{f1}.  The core radius of \objectname[]{LGS-3} is 0\farcm82 and its
optical scale length is 0\farcm78 \citep{mateo98}, so the 3\farcm4 $\times$
3\farcm4 format of the ACS covers a substantial fraction of the stars in
\objectname[]{LGS-3}.
Table \ref{logobs} gives a summary of the observations. 

\notetoeditor{Please, put Fig. \ref{f1} and Table \ref{logobs} here.}

We analyzed the images taken directly from the STScI pipeline (bias, flat-field,
and image distortion corrected). Two PSF-fitting photometry packages,
DAOPHOT/ALLFRAME \citep{ste1994} and DOLPHOT \citep{dol2000},  
were used independently to obtain the photometry of the resolved stars.
Non-stellar objects and stars with discrepant large uncertainties were 
rejected based on estimations of profile sharpness and goodness of fit. 
The final photometric lists contain $\sim 30,000$ stars each. 
See \citet{mon_etal2010b} for more details about the photometry reduction
procedures.

Individual photometry catalogs were calibrated using the equations provided 
by \citet{sir_etal2005}.  The zero-point differences between the two sets 
of photometry are $\sim 0.05$ for F475W-F814W  and $\sim 0.03$ for F814W.
These small differences are typical for obtaining HST photometry with different
methods \citep{hill98, holtzman06}.
In \S\ref{secsfhmethod} we discuss a technique which minimizes 
the already small impact of these differences in the obtained SFHs.

\begin{figure}
\centering
\includegraphics[width=14cm,angle=0]{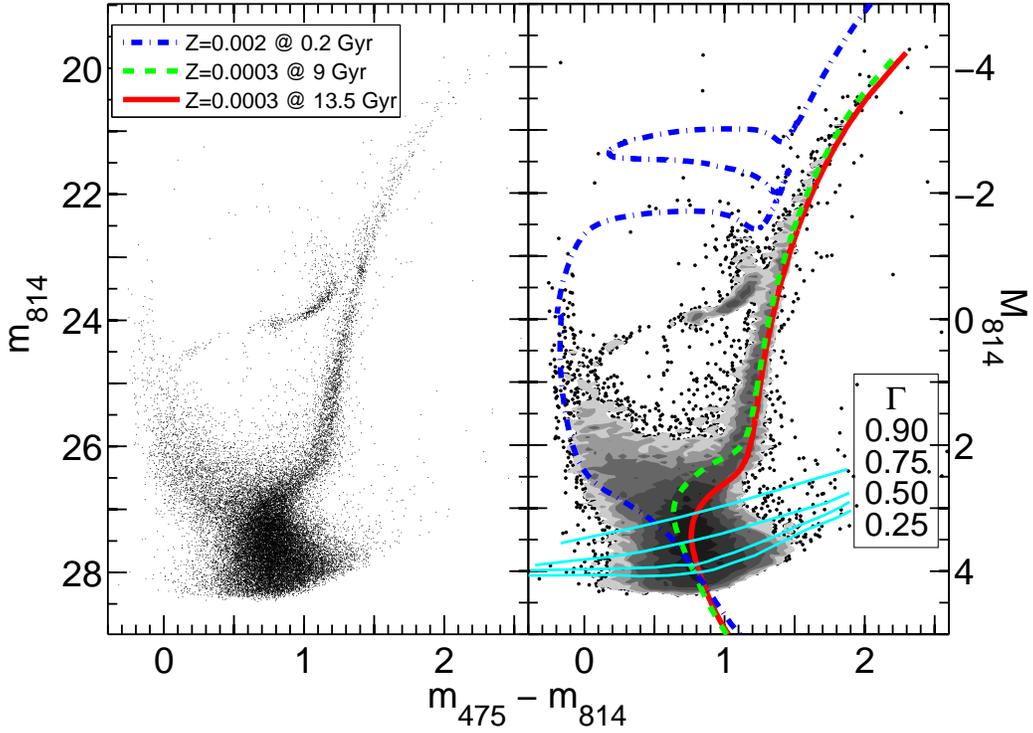}
\protect\caption[ ]{Color-magnitude diagram of \objectname[]{LGS-3}. In left panel, individual stars are plotted while right panel shows star density levels. Star densities bordering the plotted levels run from 8 to 512 stars dmag$^{-2}$, evenly spaced by factors of 2. Left axis shows magnitudes corrected from extinction. Right axis shows absolute magnitudes. A distance modulus of $(m - M)_0=24.07$ and an extinction $A_{F475w}=0.156$, $A_{F814W}=0.079$ have been used. Bottom lines in the right panel show the 0.25, 0.50, 0.75 and 0.90 completeness levels. Three isochrones from the BaSTI stellar evolution library have been over plotted. 
\label{f2}}
\end{figure}

The CMD of \objectname[]{LGS-3} is shown in Figure \ref{f2} for the DOLPHOT 
photometry. Individual stars are plotted in the left panel and density levels 
are shown in the right panel. 
The left axis shows magnitudes in the ACS photometric system corrected for extinction. 
Absolute magnitudes are given on the right axis using the adopted values for 
the distance modulus ($(m-M)_0=24.07$) and extinctions ($A_{F475w} = 0.156$ 
and $A_{F814w} = 0.079$ mag) \citep{ber2009}.
In the left panel, the lines across the bottom show the 0.25, 0.50, 0.75, and 0.90 
completeness levels derived from false star tests (discussed later in this section).
Finally, in order to highlight the main features of the CMD, 
three isochrones from the BaSTI stellar evolution library \citep{pie_etal2004} 
are also shown in the right panel.

The CMD of \objectname[]{LGS-3} shown in Figure \ref{f2} reaches $\sim 1.5$ mag 
below the oldest main-sequence turn-off; these observations are $\sim 2$ mag fainter 
than the deepest CMD previously obtained for this galaxy
\citep{mil_etal2001}. A comparison of the observations with the over-plotted 
isochrones and the presence of an extended HB indicate that a very old, very low 
metallicity stellar population is present in the galaxy. The gap produced in the 
HB by the RR-Lyrae variables is visible as well as a narrow RGB. The latter is 
populated by stars all older than 1-2 Gyr and its narrowness is an indication 
of a low spread in metallicity. The RC can be observed at the red end of the HB. 
A MS with stars younger than 200 Myr is clearly apparent. Some blue and red core 
helium-burning stars might be also present above the RC 
($m_{814}< 22.3$; $0.5<(m_{475}-m_{814})<1.2$). The RGB bump can be also observed 
at $(m_{475}-m_{814})\sim 1.4$ and $m_{814}\sim 23.2$ mag. This evolutionary 
feature may be used as indicator of a dominant coeval stellar population 
homogeneous in metallicity \citep{mon_etal2010a}.

\notetoeditor{Please, put Fig. \ref{f2} here.}

Signal-to-noise limitations, detector defects, and stellar crowding can all 
impact the quality of the photometry of resolved stars with the resulting
loss of stars, changes in measured stellar colors and magnitudes, and 
systematic uncertainties. To characterize these observational effects, we have used 
the standard technique of injecting a list of false stars in the observed 
images and obtaining their photometry in an identical manner as for the real stars. 
These observational effects must be simulated in the stars of the global, 
synthetic CMD (sCMD) (hereafter {\it synthetic} stars) to be used for the 
analysis of the SFH (see below). 

\citet{mon_etal2010b} provide a detailed description of the procedures 
we adopt for the characterization and simulation of these observational effects. 
For the present case of \objectname[]{LGS-3}, $\sim 5\times 10^5$ false stars were used for the
DOLPHOT photometry and $\sim 6.5\times 10^5$ were used for the DAOPHOT 
photometry. The completeness factor, $\Gamma$, is defined as the rate of 
recovered to injected stars as a function of magnitude and color. 
Figure \ref{f2}, right panel, shows $\Gamma=0.25, 0.50, 0.75$ 
and 0.90 for the DOLPHOT photometry. Results for DAOPHOT are similar. 
At the magnitude level of the oldest MS turn-off stars, $\Gamma\sim 0.90$,
allowing for very strong constraints on the oldest epochs of star formation. 

We simulate the observational effects in the stars of the sCMD
with the routine {\it obsersin}. 
It uses the unrecovered false stars and the 
differences between the injected ($m_i$) and measured magnitudes ($m_r$) of 
the recovered stars for the simulation \citep{apa_gal1995}. 
Since the density of stars (and thus the amount of crowding) normally varies 
substantially across the galaxy's field, spatial information has also been 
taken into account in the completeness test. To do so, first the spatial density distribution of 
observed stars in \objectname[]{LGS-3} is measured. For that, a grid of $100\times 100$ pixel 
size bins is defined, stars are counted in each bin and the result is smoothed 
and normalized. This is used as a density probability distribution to assign 
positions to each synthetic star. Then, for each synthetic star with 
magnitude $m_s$, a list of the false stars with $|m_i - m_s|\leq \epsilon_m$; 
$r\leq \epsilon_r$ is created for each filter, where $r$ is the spatial distance 
between synthetic and false star and $\epsilon_m$ and $\epsilon_r$ free input 
searching intervals. If the number of false stars selected in this way, in 
common to both filters, is $<10$, the parameters $\epsilon_m$ and $\epsilon_r$ 
are increased in turn until 10 or more false stars are found. Then, one of 
them is randomly selected from this list. If that star was unrecovered, 
the synthetic star is eliminated from the sCMD. If the selected star was 
recovered and $m_i^\prime$ and $m_r^\prime$ are its injected and recovered 
magnitudes in a given filter, then $m_s^e=m_s + m_i^\prime- m_r^\prime$ will 
be the magnitude of the synthetic star with observational effects simulated. 
The same is done for both filters.

\section{The IAC Method to Solve for the SFH}\label{secsfhmethod}

\subsection{Details of the IAC-star, IAC-pop, and MinnIAC codes}

Our analysis of the SFH of \objectname[]{LGS-3} and the rest of LCID galaxies is done upon two fundamental requirements: i) the consistency of the method, in order to obtain SFHs that are readily comparable and ii) the stability and uniqueness of the solution. To accomplish i), we have adopted the same criterion on the selection of input parameters to derive the SFH for all galaxies of the LCID project. For ii), we have further developed our method and created new tests and control procedures. Details are given in the following and appendix \ref{secimf}. 

\begin{figure}
\centering
\includegraphics[width=14cm,angle=0]{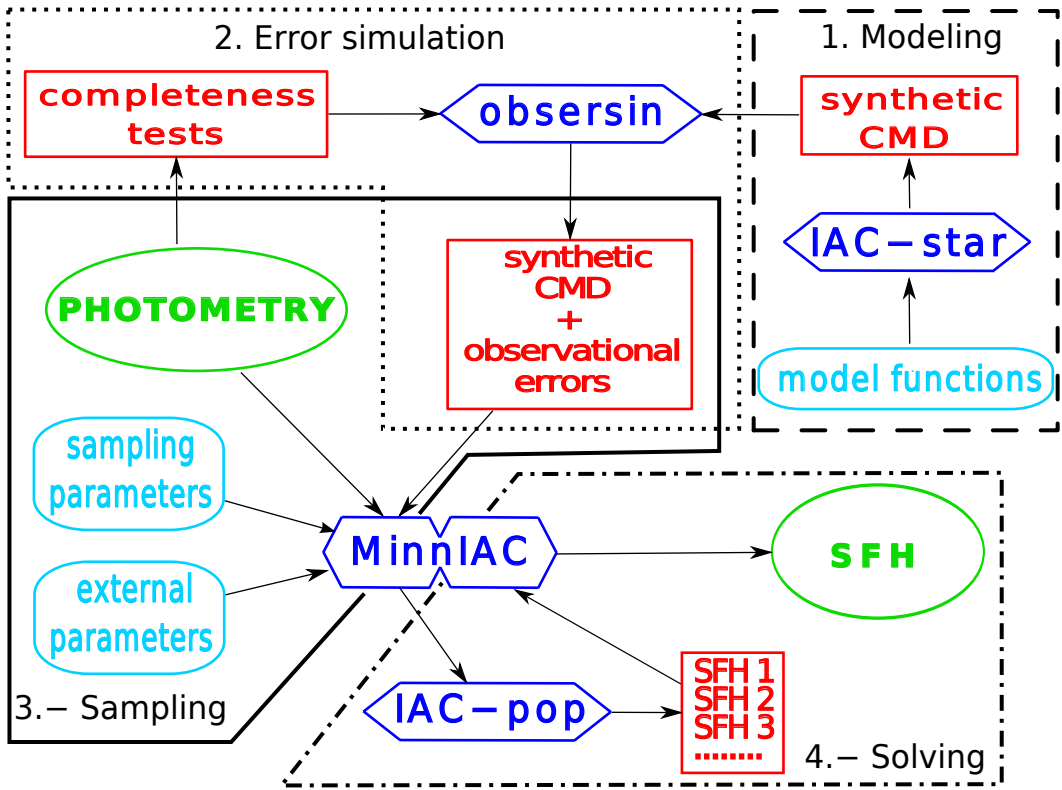}
\protect\caption[ ]{Data flow diagram followed in the LCID project to obtain the SFHs of the galaxies. See text for details. 
\label{f3}}
\end{figure}

The IAC method to solve the SFH is based in three main codes: (i) IAC-star \citep{apa_gal2004} is used for the computation of synthetic stellar populations and CMDs; (ii) IAC-pop \citep{apa_hid2009} is the core algorithm for the computation of SFH solutions, and (iii) MinnIAC \citep[introduced in][but explained in more detail in this paper]{apa_hid2009} is a suite of routines created to manage the process of sampling the parameter space, creating input data, and averaging solutions. As a summary, Figure \ref{f3} shows a data flow diagram for the SFH computation process.

\notetoeditor{Please, put Fig. \ref{f3} here.}

Considering that time and metallicity are the most important variables in the problem \citep[see][]{apa_hid2009}, we define the SFH as an explicit function of both. We will use two definitions, one in terms of number of formed stars and another in terms of gas mass converted into stars. Formally, $\psi_N(t,Z){\rm d}t{\rm d}Z$ is the number of stars formed within the time interval $[t,t+{\rm d}t]$ and within the metallicity interval $[Z,Z+{\rm d}Z]$, while $\psi(t,Z){\rm d}t{\rm d}Z$ is the mass converted into stars within the same intervals. The relation between $\psi_N(t,Z)$ and $\psi(t,Z)$ is the average stellar mass at birth. Both can be identified with the usual definitions of the SFR, but as a function of time and metallicity. 

Several criteria can be used to grid the CMD. The simplest is to use a uniform grid and count stars within each box. However, not all the regions of the CMD and stellar evolutionary phases contain information of the same quality for the purpose of the SFH derivation. An alternative is to give higher weights to regions with more certain stellar evolutionary phases (like the main sequence or the sub-giant branch),
regions which are more heavily populated, or regions in which the photometry is more accurate. 
To do so, we have introduced what we call {\it bundles} \citep[see][and Figure \ref{f4}]{apa_hid2009}, that is, macro-regions on the CMD that can be subdivided into boxes using different appropriate samplings. The number and sizes of boxes in each bundle determines the weight of that region in the derived SFH. 
This is an efficient and flexible approach for two reasons. First, it allows a finer sampling in those regions of the CMD where the models are less affected by uncertainties in the input physics. Second, the box size can be increased, and the impact on the solution decreased, in those regions where the small number of observed stars could introduce noise due to small number statistics or where we are less confident in the stellar evolution predictions.

IAC-pop provides a solution for the input set of data and parameters. 
It can also provide several solutions for the same set by changing the initial 
random number generator seed. Alternatively, it can deliver several 
solutions by randomly changing the input data $O^j$ (the number of stars 
in the original, oCMD, boxes) according to a Poisson statistics. 
However all of these solutions are eventually sensitive to several 
other input functions and parameters. 
To better analyze their effects, we can divide them into four groups. 
The first group consists of {\it sampling parameters}; the main ones are 
the number and boundaries of parent, pCMDs, and the criteria to define the bundles 
and grid arrays for sampling the CMDs. 
The second group consists of {\it external parameters}; these are the photometric 
zero points, distance modulus, and reddening. 
The third group consists of {\it model functions}, which are those functions 
defining the sCMD; the main ones are the initial mass function (IMF), $\phi(m)$, 
and the function accounting for the frequency and relative mass distribution 
of binary stars, $\beta(f,q)$ \citep[see][]{apa_hid2009}. Finally, we have
the {\it external libraries}, which consists of stellar evolution and bolometric 
correction libraries. Limiting and measuring the effects of all these 
parameters in the SFH solution requires several thousand runs of IAC-pop 
using different input sets. MinnIAC is used to test for the effects of the 
four groups are treated separately. In general, the solutions obtained with 
varying {\it sampling parameters} are averaged; their rms deviation is also a 
good estimate of the solution internal error. As for the {\it external parameters}
and {\it model functions}, the set with the best $\chi_\nu^2$ is usually adopted. 
Regarding the {\it external libraries}, the solution with the best $\chi_\nu^2$ 
may be adopted, but all the solutions are usually conserved; their differences 
can provide an indication of the external uncertainties associated to such 
libraries.

MinnIAC (from {\it Minn}esota and {\it IAC}) is a suite of routines developed specifically to manage the process of selection of sampling parameters, creating input data sets, and averaging solutions. More in detail, regarding the parameter selection and data input, MinnIAC is used for two tasks. First, to divide the sCMD in the corresponding pCMD according to the selected age and metallicity bins. Second, to define the bundles and box grids in the CMDs and counting stars in them. With this information, MinnIAC creates the input files for each IAC-pop run. Several selections of both simple populations and bundle grids are used within each SFH solution process. MinnIAC does this task automatically, requiring only the input of bundles and corresponding grid sizes, age and metallicity bin sizes and the steps by which the starting points of the CMD grid as well as age and metallicity bins should be modified from one IAC-pop run to another. A second module of MinnIAC is used to average single solutions found by IAC-pop as required. 

\subsection{Solving procedure, results, and uncertainties for LGS-3}

In this section we will describe with some details the specific application of the IAC method to the case of \objectname[]{LGS-3}, detailing the different effects mentioned in \S\ref{secsfhmethod}: {\it sampling parameters}, {\it external parameters}, {\it model functions} and {\it external libraries}. Our choices are summarized in Table \ref{varpar}. For each item, Columns 2, 3, and 4 give minimum (or choice 1), maximum (or choice 2) and the step used (if relevant), respectively. Column 5 gives the final adopted value (if relevant). The SFH of \objectname[]{LGS-3} has been derived independently for the DOLPHOT and DAOPHOT photometry sets following the same procedure as described below. 

Using IAC-star, several sCMDs were computed, one with each of the IMFs and binarity functions listed in Table \ref{varpar}. Each sCMD was built with $8\times 10^6$ stars with a constant star formation rate in the range of age 0--13.5 Gyr and a metallicity range 0.0001--0.005. The latter value has been assumed after a preliminary test with isochrones which showed that stars with metallicity larger than 0.005 do not exist in \objectname[]{LGS-3}. Using a large number of synthetic stars in the sCMD is important in order to minimize the stochastic effects of the Monte Carlo simulations used in IAC-star; to have a statistically significant number of stars in each pCMD ($>6000$ stars), and to reduce the uncertainties in the stellar composition of simple populations in the pCMDs.

\begin{figure}
\centering
\includegraphics[width=14cm,angle=0]{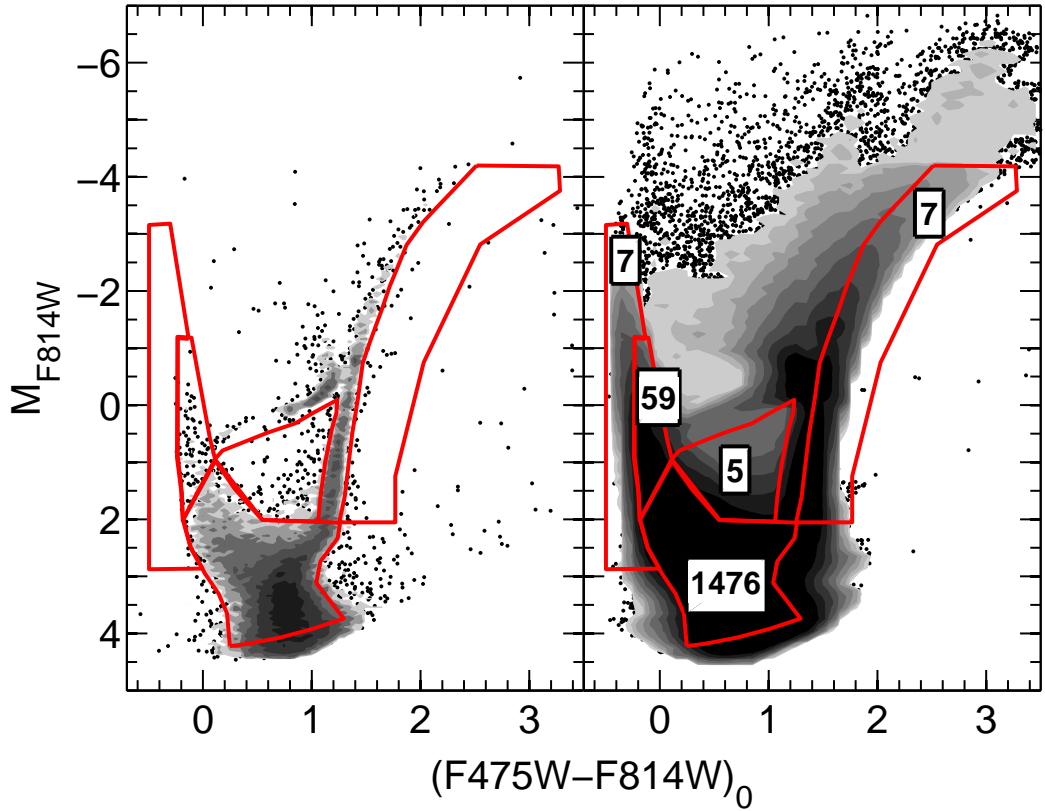}
\protect\caption[ ]{Regions of the CMDs (bundles) used for gridding. Bundles are shown overplotted on the observed CMD (left panel) and on one of the synthetic CMD used as an input model for the SFH derivation (right panel). Quantities in right panel give the number of bins used to sample each bundle.\label{f4}}
\end{figure}

The bundles and one of the grid sets used to bin the CMDs are shown in Figure.\ref{f4}. In total, 72 combinations of IMF and binarity ({\it model functions}) have been tested. For each of them, 24 solutions have been obtained by varying the CMD binning and the simple stellar population sampling ({\it sampling parameters}). The average $\bar\psi$ of these 24 solutions is adopted as the solution for the specific IMF and binarity. These averages have been obtained by using a boxcar of width 1 Gyr and step 0.1 Gyr for $t$ and width $0.0006$ and step $0.0001$ for $Z$. A $\bar{\chi^2_\nu}$ is also obtained as the average of the 24 single $\chi^2_\nu$ values. The average solution with the smallest $\bar{\chi^2_\nu}$ is assumed as the best one. The IMF and the binary fraction corresponding to this solution are assumed here as the ones best representing our photometry data, although we do not intend to reach any strong conclusions concerning their true values. 

\notetoeditor{Please, put Fig. \ref{f4} and Table \ref{varpar} here.}

To minimize the effects of {\it external parameters} (distance, reddening and photometry zero-point), the SFH is derived for different offsets in color and magnitude applied to the oCMD. The offset giving the minimum $\bar{\chi^2_\nu}$ is assumed as the best. These are obtained when the oCMD are shifted by $[\Delta($F475W$-$F814W$),\Delta($F814W$)]=[0.02,0.11]$ and $[-0.03,0.11]$ for the DOLPHOT and DAOPHOT photometry sets, respectively. These tests also provide a way to estimate how distance, reddening and zero-point uncertainties translate into the solution of the SFH. 
Defining $\bar{\chi^2}_{\nu,min}$ as the $\bar{\chi^2_\nu}$ value of the best solution, if $\sigma_{\bar{\chi^2}_{\nu,min}}$ is its corresponding rms deviation, all the solutions with $\bar{\chi^2}_\nu\leq\bar{\chi^2}_{\nu,min}+\sigma_{\bar{\chi^2}_{\nu,min}}$ should be considered valid. We can assume that the rms deviation of all the corresponding single solutions is a proxy of the error of the adopted solution, $\sigma_P$.

The former error, $\sigma_P$, includes the effects of {\it sampling} and {\it external parameters}. However, there is one more source of uncertainty that must be taken into account, namely the effect of statistical sampling in the oCMD. To estimate this, the solution is again obtained after varying the numbers of stars in the oCMD grid according to a Poisson statistic. This is done 20 times and the rms deviation of the solutions, $\sigma_I$, is obtained. The final adopted error is the sum in quadrature $\sigma_{SFH}=(\sigma_I^2+\sigma_P^2)^{1/2}$.

Finally, solutions have been obtained using both the BaSTI \citep{pie_etal2004} and the \citet{gir_etal2000} stellar evolutionary libraries ({\it external libraries}). In both cases the bolometric corrections by \citet{bed_etal2005} were used. 
Both produced qualitatively similar results and a comparison between them together with the results obtained with other methods is presented in \ref{comresult}. 
For detailed analysis presented in the following sections, for 
simplicity, we use only the BASTI stellar evolutionary libraries.

\section{The SFH of LGS-3}\label{secsfhresult}

\subsection{Main Features of the SFH of LGS-3}\label{secsfhfeatures}

\begin{figure}
\centering
\includegraphics[width=12cm,angle=0]{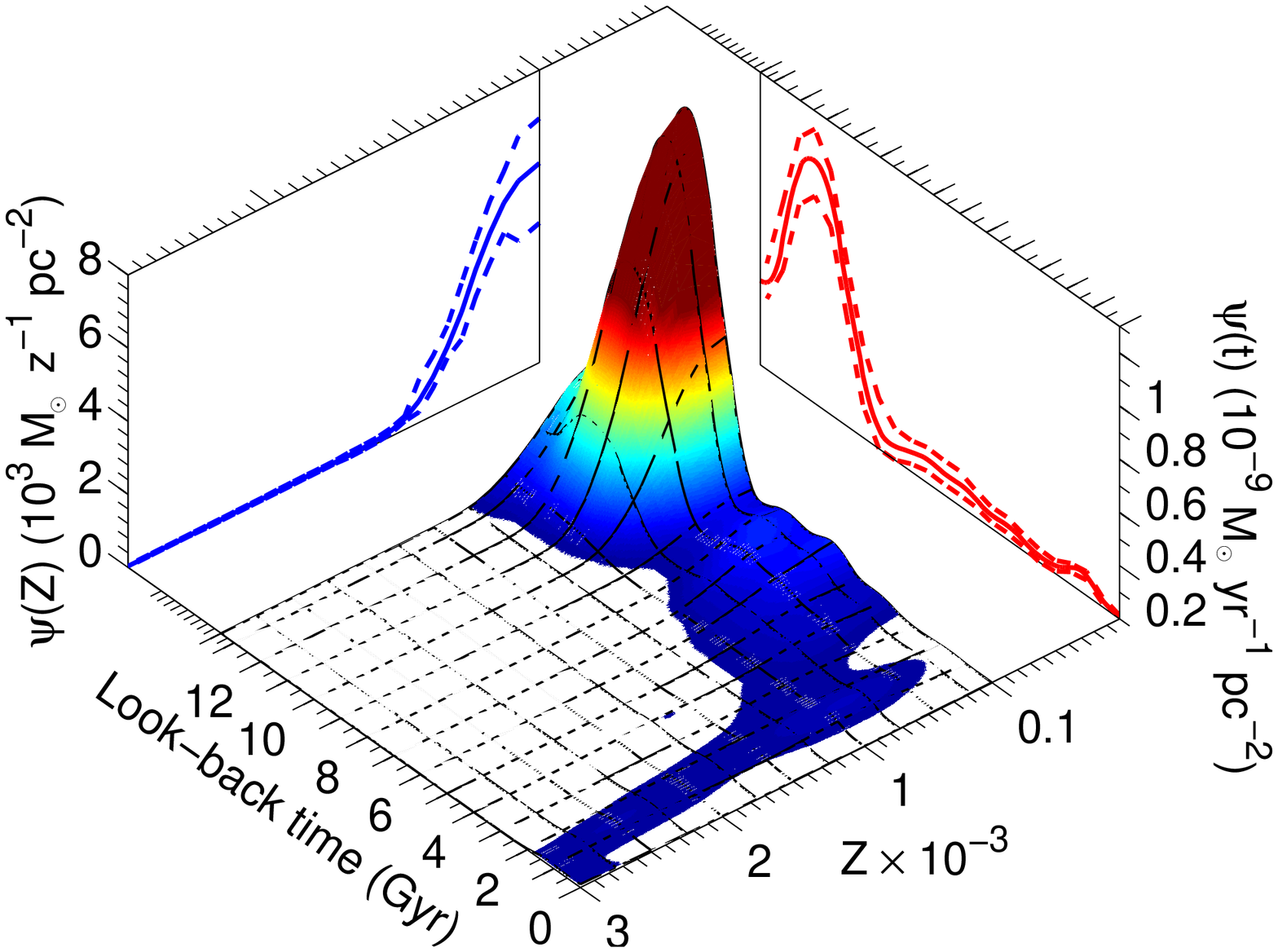}
\includegraphics[width=12cm,angle=0]{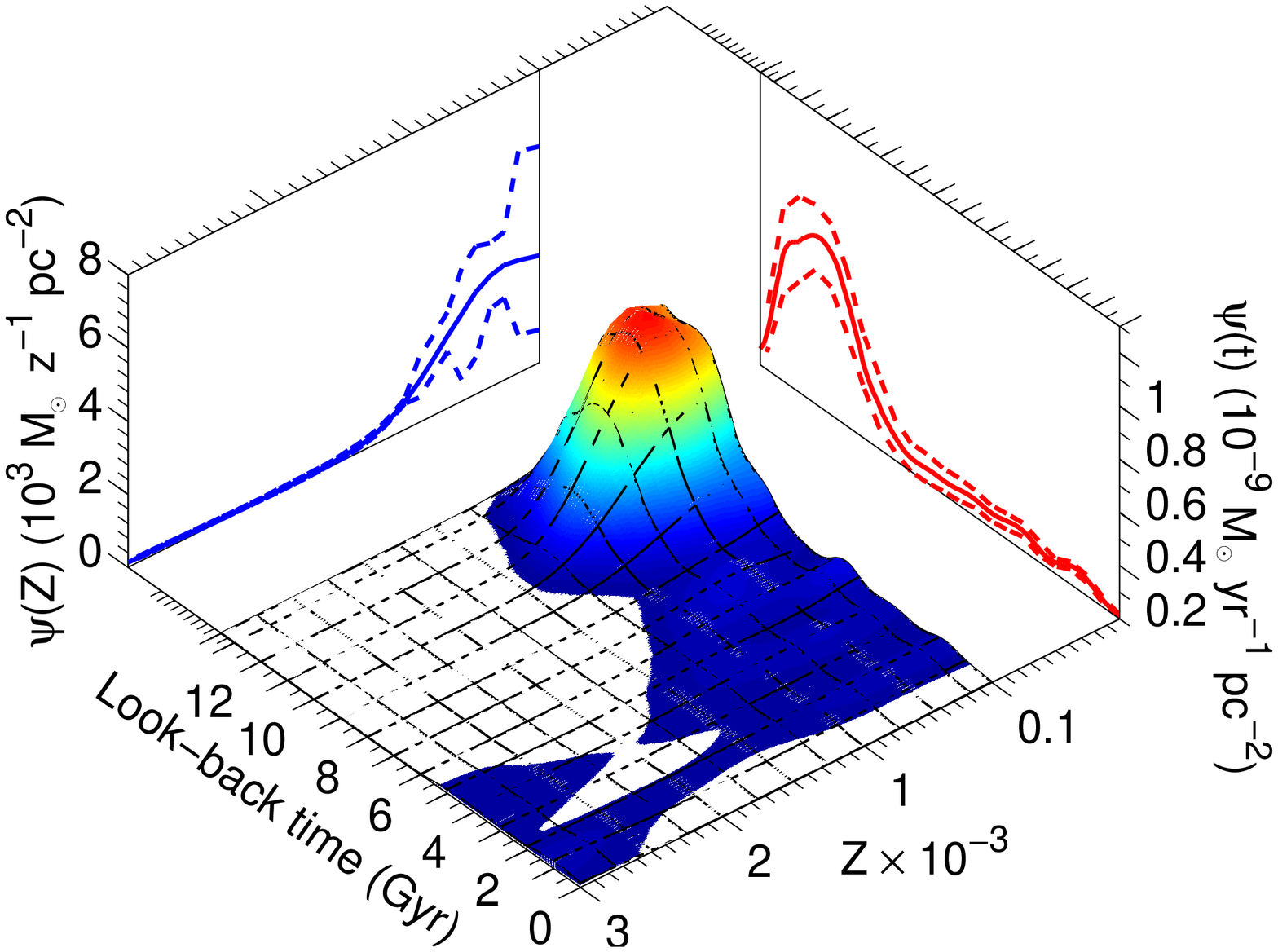}
\protect\caption[ ]{Star formation history, $\psi(Z,t)$, of \objectname[]{LGS-3} obtained from DOLPHOT (upper panel) and DAOPHOT (lower panel) photometry sets. The SFH as a function of age $\psi(t)$ and metallicity $\psi(Z)$ are shown projected on the $\psi-time$ and $\psi-metallicity$ planes, respectively. Dashed lines give the error intervals. The age-metallicity relationship is the projection onto the look-back time--metallicity plane.
\label{f5f6}}
\end{figure}

Figure \ref{f5f6} shows 
plots of the SFH of \objectname[]{LGS-3}, $\psi(t,Z)$, as a function of both $t$ and $Z$
derived from the two different photometry sets.
$\psi(t)$ and $\psi(Z)$ are also shown 
in the $\psi-t$ and $\psi-Z$ planes, respectively. The projection of $\psi(t,Z)$ on 
the age-metallicity plane shows the AMR,  including the metallicity dispersion. 

\notetoeditor{Please, put Fig. \ref{f5f6} here.}

\begin{figure}
\centering
\includegraphics[width=14cm,angle=0]{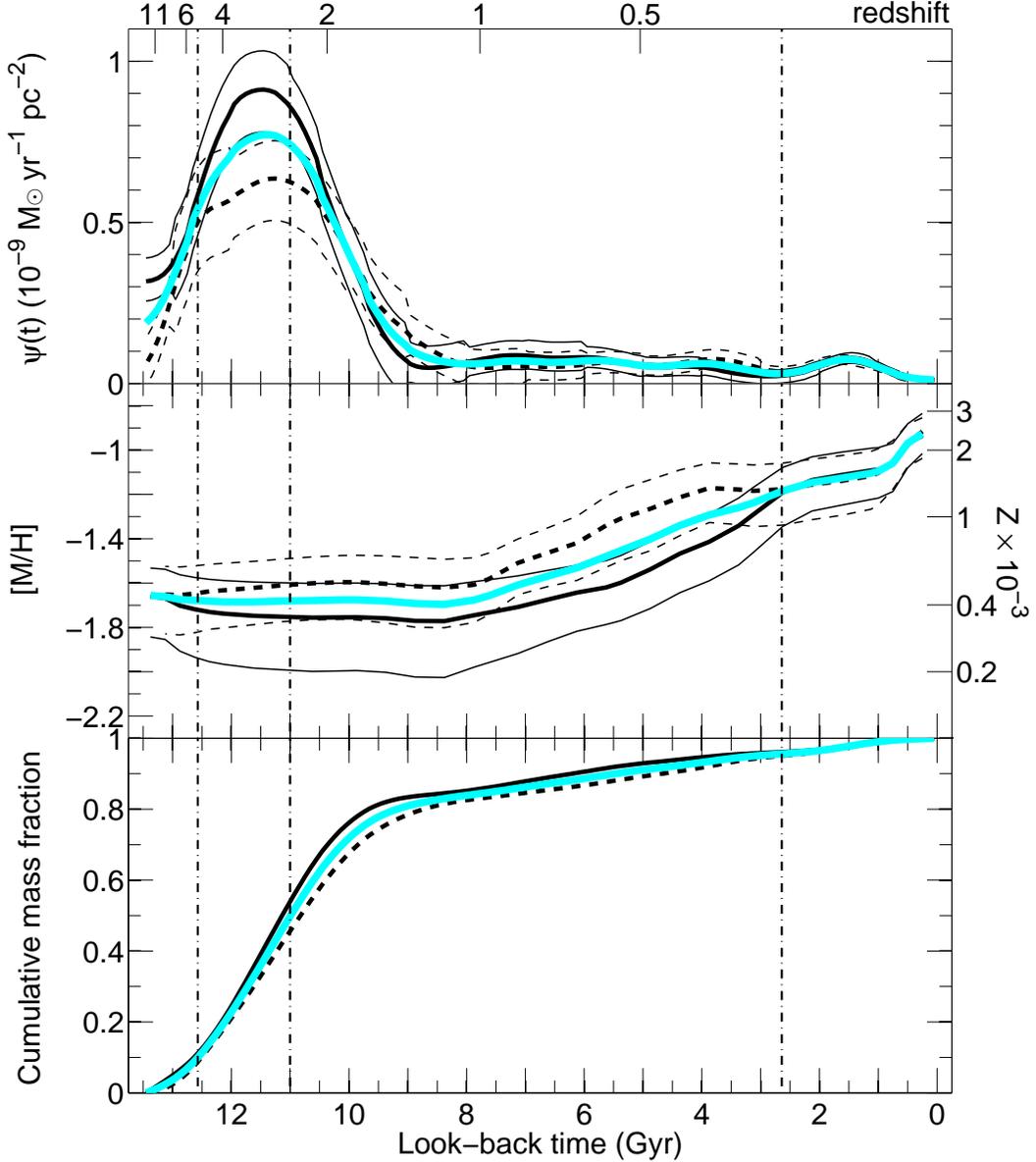}
\protect\caption[ ]{$\psi(t)$ (upper panel), metallicity (middle panel), and cumulative mass fraction (lower panel) of \objectname[]{LGS-3} obtained from DOLPHOT (thick, solid lines) and DAOPHOT (thick, dashed lines) photometry sets. Thin lines give the uncertainties. Cyan lines show the final adopted solution, which is the mean of DOLPHOT and DAOPHOT ones. Vertical dotted-dashed lines indicate the times corresponding to the 10$^{\rm th}$, 50$^{\rm th}$ and 95$^{\rm th}$ percentiles of $\psi(t)$, i.e. the times for which the cumulative fraction of mass converted into stars was 0.1, 0.5 and 0.95 of the current one. A redshift scale is given in the upper axis, computed assuming $H_0=70.5\rm~km~s^{-1}~Mpc^{-1}$, $\Omega_m=0.273$, and a flat Universe with $\Omega_\Lambda = 1 - \Omega_m$. 
\label{f7}}
\end{figure}

In Figure \ref{f7}
we show $\psi(t)$, the AMR, and the cumulative mass fraction of \objectname[]{LGS-3}, as a 
function of time with their associated errors. Three vertical dashed lines show 
the ages of the 10th, 50th, and 95th-percentile of the integral of $\psi(t)$. 
Table \ref{meansolpar} gives a summary 
of the integrated and mean quantities for the observed area and estimates for
the whole galaxy.

The main features of the SFH of \objectname[]{LGS-3} can be observed in Figure 
\ref{f7}. 
The bulk of the star formation of \objectname[]{LGS-3} occurred at old epochs, 
with a main peak located $\sim 11.5$ Gyr ago. 
The initial episode of SF lasted until roughly 9 Gyr ago. 
Approximately 85\% of the stars were formed in this initial episode. 
Since then, the star formation rate has been very low and slowly declining 
until the present. 
The SFR at the measured main peak is a factor of 15 larger than the mean 
SFR between 0 and 9 Gyr ago (because of limited time resolution, 
this is a lower limit, see \S\ref{secconstraints}). 
It is worth noting that the age at which the rate of the AMR starts to increase more steeply matches up with the age at which the SFR decreases to a low level. 
This may indicate a scenario in which the stars were formed at a higher rate at old epochs from fresh, low metallicity gas. 
When the initial episode of star formation was over, the gas reservoir would have been 
nearly exhausted. As a consequence, subsequent star formation would occur at a much 
lower rate and from an ISM including a high fraction of processed gas, 
hence significantly increasing the metallicity enrichment rate. 

\notetoeditor{Please, put Fig. \ref{f7} and Table \ref{meansolpar} here.}

It is interesting to test whether the number of SNe produced in \objectname[]{LGS-3} during 
its lifetime are sufficient to produce the chemical enrichment we have derived. 
Focusing on the main episode, $1.7\times 10^6$ M$_\odot$ of gas were converted 
into stars from 13.5 to 8.5 Gyr ago. Assuming a minimum progenitor mass for core 
collapse SNe of 6.5 M$_\odot$ \citep{sal_cas2005} we obtain a total of 
$2.9\times 10^4$ SNe produced in the old episode.  Accounting for stellar mass loss, 
and assuming a minimum initial mass of 10 M$_\odot$ for SNe progenitors, then the 
total number of SNe decreases to $1.6\times 10^4$. This can be compared to an
estimate of the total number of SNe necessary to produce the observed chemical 
enrichment. Assuming the solar abundance distribution from \citet{gre_noe1993},  
that only SNe II are involved, and using the iron mass production rate from 
\citet{nom_etal1997} \citep[see][for more details]{mon_etal2010c}, we estimate 
that $\sim$ 80 SNe II are enough to produce an increment of metallicity from 
0 to 0.0007, which is the upper value obtained for \objectname[]{LGS-3} 8.5 Gyr ago. 
This estimate was derived assuming the iron mass production of a 13 $M_{\odot}$ star 
(which may vary by a factor of 2--3 with different assumptions). 
Although relatively uncertain, our calculation shows that the galaxy needs to 
retain only roughly 1\% of the material released by SNe in order to produce the 
observed metallicity increase. The same calculation can be made for the chemical 
evolution from 8.5 Gyr ago to the present. In this case, the mass of gas converted 
into stars is $2.4\times 10^5$ M$_\odot$ and the corresponding numbers of SNe are
$4.1\times 10^3$ and $2.4\times 10^3$.  In this time, the metallicity increases 
from $Z=0.0007$ to a current upper value of $Z=0.003$, requiring 290 SNe. This
is less than 15\% of the total, a much larger fraction than calculated for the
initial star formation episode.  This implies that a significantly larger fraction
of the newly synthesized material was captured at later times at lower SFRs.
Note that only SNe II have been considered for this estimates; if SNe Ia were 
also included, even lower fractions of captured newly synthesized material would result.  Note also, that much of the latter enrichment could be achieved by
SNe Ia from the initial star formation episode.

The SFH shown in Figure \ref{f7} 
corresponds to the observed area, which covers the center and a major fraction 
of the galaxy.  The total mass of stars created in this area over the lifetime
of the galaxy is $2.0\times 10^6$ M$_\odot$. 
Assuming an exponential density distribution of scale-length 112 pc (S.\ Hidalgo
et al., in prep.), we estimate that the total mass of stars ever formed in \objectname[]{LGS-3} 
is $1.1\times 10^6$ M$_\odot$. 
Adding in the gas mass \citep[][see \S\ref{secint}]{you_lo1997}, then the 
total baryonic mass of \objectname[]{LGS-3} is $1.5\times 10^6$ M$_\odot$. 
These values correspond to the best solution with the adopted IMF ($x=1.3$ for the interval $0.1\leq m\leq 0.5$ M$_\odot$; $x=2.3$ for the interval $0.5\leq m\leq 100$ M$_\odot$; see Table \ref{varpar}). 
If a Salpeter IMF with an exponent of 2.35 for the range 0.1 to 100 M$_\odot$ is used, 
then the result would be larger by a factor of $\sim 1.5$.

\begin{figure}
\centering
\includegraphics[width=14cm,angle=0]{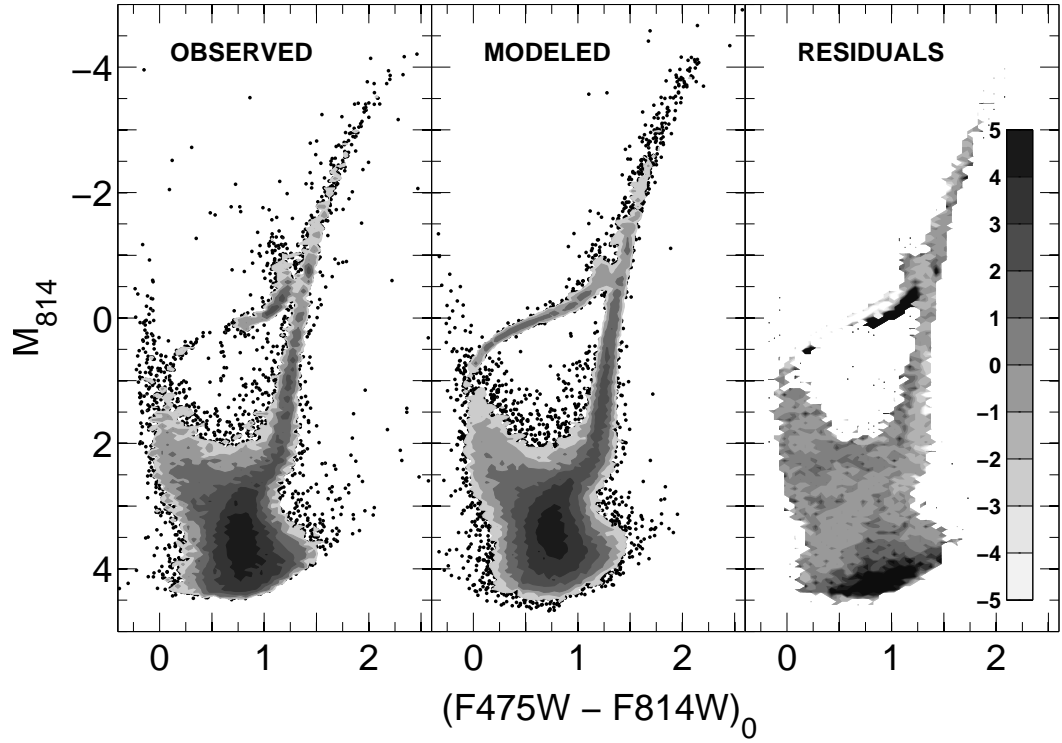}
\protect\caption[ ]{Observed (left panel), calculated (central panel), and residuals (right panel) CMDs. The calculated CMD has been built using IAC-star with the solution SFH of \objectname[]{LGS-3} as input. The grey scale and dot criteria for the first two is the same as for Figure \ref{f2}. The residuals are in units of Poisson uncertainties.\label{f8}}
\end{figure}

Figure \ref{f8} shows the observed CMD, the best-fit CMD, and the corresponding residuals. The calculated CMD has been built using IAC-star with the solution SFH of \objectname[]{LGS-3} as input. Residuals are given in units of Poisson uncertainties, obtained as $(n^o_i-n^c_i)/\sqrt{n^c_i}$, where $n^o_i$ and $n^c_i$ are the number of stars in bin $i$ of an uniform grid defined on the observed and calculated CMDs, respectively. 

\notetoeditor{Please, put Fig. \ref{f8} here.}

\subsection{Estimating the effects of time resolution on the SFH}\label{secconstraints}

\begin{figure}
\centering
\includegraphics[width=14cm,angle=0]{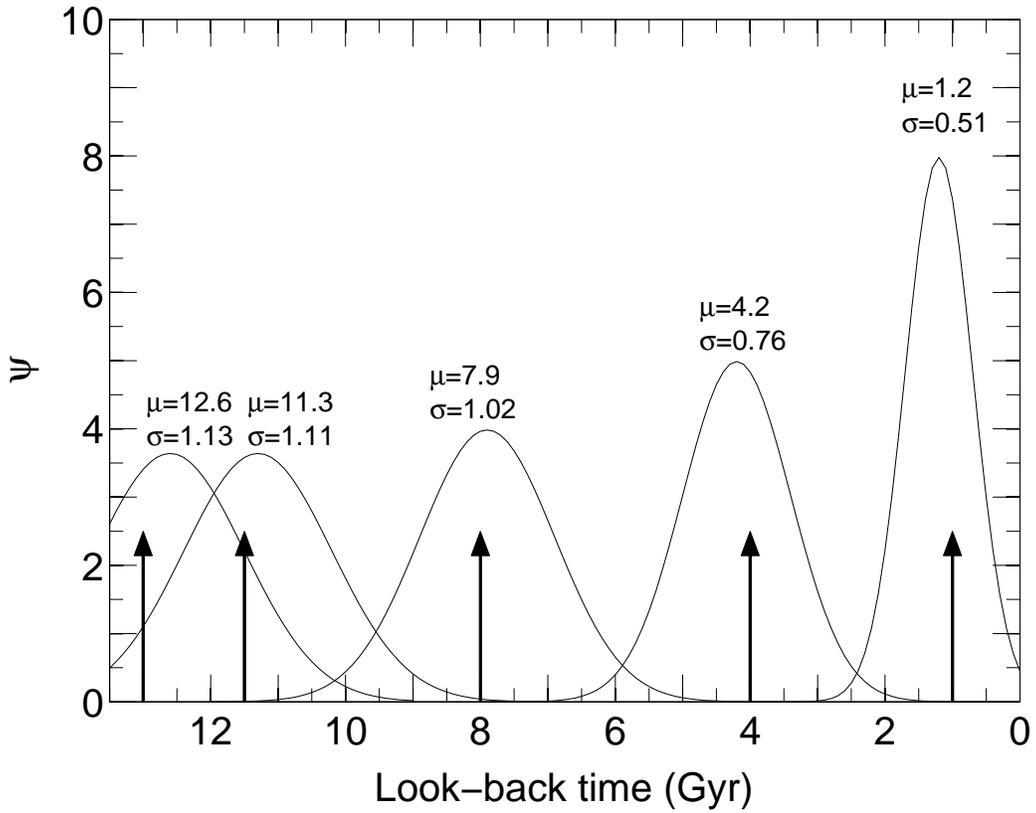}
\protect\caption[ ]{Results of the tests of time resolution. Arrows indicate the burst age of the simulated populations. Bursts have a duration of $10^5$ M$_\odot$ and their intensities are similar to the SFR of \objectname[]{LGS-3} at the corresponding age. Arrows heights are not indicative of burst intensity. Curves show the Gaussian profile fits to the solutions. The Gaussian profile areas are normalized to the same value for easier comparison. The solution peaks ($\mu$) and standard deviations ($\sigma$) are given.
\label{f9}}
\end{figure}

Observational uncertainties as well as the uncertainties inherent to the SFH computation 
procedure result in a smoothing of the derived SFH and a decrease of the age resolution 
(see appendix \ref{stab}). Having a accurate estimate of this age resolution is 
fundamental in order to obtain reliable information of the actual properties of the 
galaxy. We have calculated the age resolution as a function of time by recovering the 
SFH of a set of mock stellar populations corresponding to very narrow bursts occurring
at different times. Five mock populations have been computed, each one with a single 
burst at ages of 1, 4, 8, 11.5, and 13 Gyr. We used IAC-star to create the CMDs 
associated with each mock stellar population. For the mock populations at ages of 11.5 
and 13 Gyr, five simulations have been done with different random number seeds in order 
to better characterize them. The assumed metallicities and SFRs correspond to those obtained 
for \objectname[]{LGS-3}. Observational uncertainties were simulated in the CMDs by using the results 
of the DOLPHOT completeness tests.  The SFHs of the mock stellar populations were 
then obtained using the identical procedure as for the real data. In all cases, 
the recovered SFH is well fitted by a Gaussian profile with $\sigma$ depending on 
the burst age. Figure \ref{f9} shows the input bursts and the 
corresponding Gaussian profile fits to the solutions, including their peak age and 
$\sigma$. From these results we estimate an intrinsic age resolution of 
$\sim 1.1$ Gyr at the oldest ages which decreases to $\sim 0.5$ Gyr at an age of 1 Gyr.
Also, a slight shift is observed between the input and solution ages. For the oldest trials,
the shift is toward younger ages (up to $~0.4$ Gyr for the oldest trial) and for the
youngest trials it is toward older ages (about 0.2 Gyr).

\notetoeditor{Please, put Fig. \ref{f9} here.}

It is particularly important to determine the oldest age at which star formation 
occurred in \objectname[]{LGS-3} and the real duration of the first, main star formation episode. 
Our SFH solution (Figure \ref{f7}) 
is well fitted by a Gaussian profile with a peak, $\mu_{obs}=11.5$ Gyr 
and $\sigma_{obs}=1.2$ Gyr in the age range 9.5--13.5 Gyr (using the DOLPHOT photometry). 
The shifts shown in Figure \ref{f9} indicate that the actual 
star formation peak would have occurred at about 11.7 Gyr ago. A comparison of
the solution with our estimated resolution indicates that the initial episode
of star formation is just barely resolved in our observations.  A 
quadratic subtraction of the $\sigma\sim 1.1$ that we have obtained for a 
sharp burst at this age from the derived $\sigma_{obs}=1.2$ provides a first estimate of 
the actual, intrinsic dispersion of $\sigma_{real}\sim 0.5$. 

\begin{figure}
\centering
\includegraphics[width=14cm,angle=0]{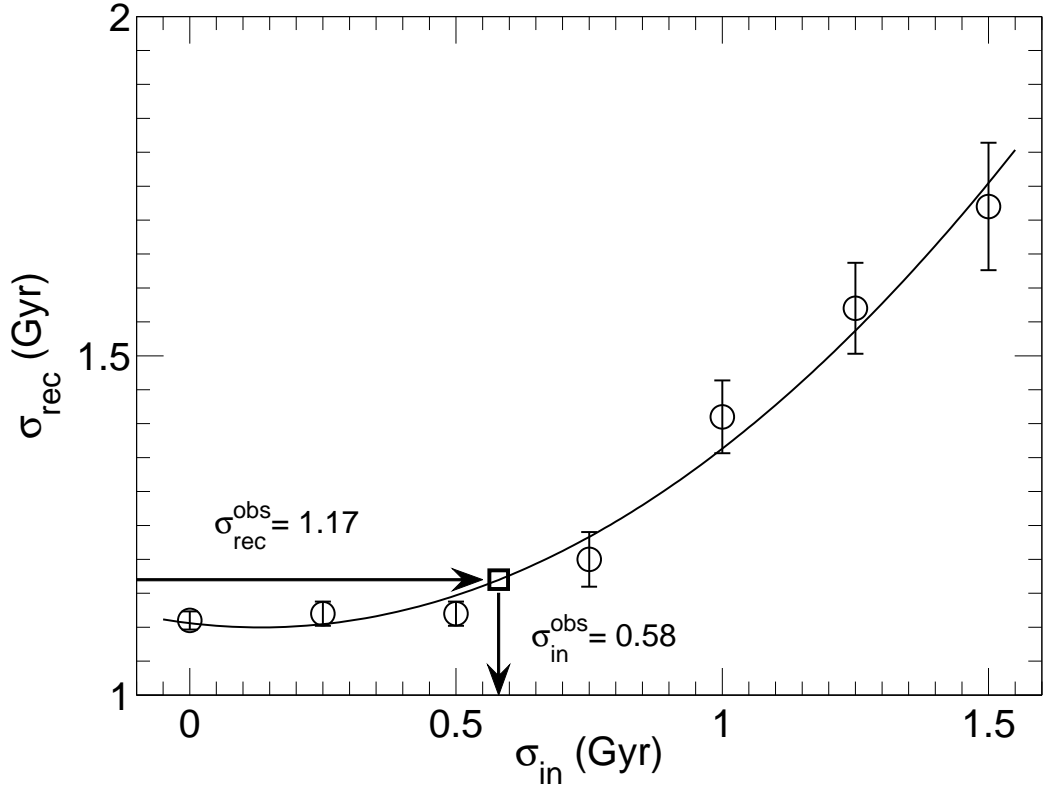}
\protect\caption[ ]{The standard deviation, $\sigma^{mock}_{rec}$, of Gaussian profile fit solutions obtained for a set of Gaussian profile mock input SFHs with standard deviation $\sigma^{mock}_{in}$, centered at about the age of the peak of the main episode of \objectname[]{LGS-3} (11.5 Gyr). Observational uncertainties were simulated in the associated mock CMDs and the SFHs were recovered following the same procedure as for \objectname[]{LGS-3}. A quadratic fit to the data is shown, together with the $\sigma$ value measured in the real data solution ($\sigma_{\rm rec}^{\rm obs}$) and the corresponding real value inferred using the fit ($\sigma_{\rm in}^{\rm obs}$).
\label{f10}}
\end{figure}

Given the importance of this measurement, we have pursued a more detailed analysis. 
We have extended the former tests to a set of mock stellar populations with SFHs 
defined as Gaussian profile shaped bursts of different $\sigma$ and centered 
at 11.5 Gyr. Observational uncertainties were simulated in the associated CMDs 
and the SFHs were recovered following the same process as for the galaxy.
In all cases, the recovered SFHs are well fitted by Gaussian profiles in the 
age range 9.5--13.5 Gyr. Figure \ref{f10} shows a quadratic 
fit to the standard deviation of the solutions, $\sigma^{mock}_{rec}$, as a 
function of the input ones, $\sigma^{mock}_{in}$. If the actual \objectname[]{LGS-3} SFH were 
well described as a Gaussian profile, its $\sigma$ could be inferred by 
interpolating in this fit, which gives $\sigma^{obs}_{in}=0.6$ Gyr. We will 
adopt this value for the forthcoming discussion. 

\notetoeditor{Please, put Fig. \ref{f10} here.}

\subsection{Comparison with other SFHs derivation methods.}\label{comresult}

As for the rest of the LCID galaxies, we have obtained the SFH of \objectname[]{LGS-3} using 
the Match \citep{dol2002} and Cole \citep{ski_etal2003} methods in addition 
to the IAC method. These two methods use \citet{gir_etal2000} for the stellar 
evolution library and we have adopted the DOLPHOT photometry to be used with 
them for this comparison \citep[for a description of the main features of 
these methods and the particulars of their application to LCID 
galaxies, see][]{mon_etal2010b}. For a proper comparison, we have also 
obtained the SFH with IAC-star/MinnIAC/IAC-pop using the same library and 
photometry. 
This also allows us to compare the solutions obtained with the IAC method 
using two different stellar evolution libraries.

\begin{figure}
\centering
\includegraphics[width=14cm,angle=0]{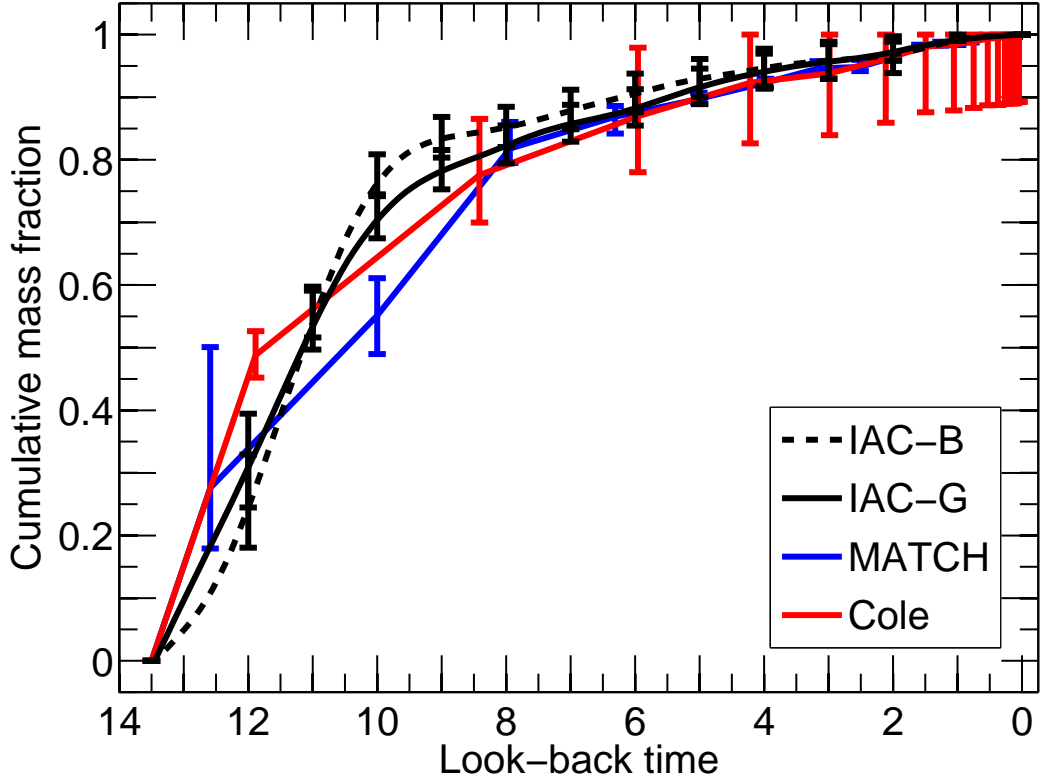}
\protect\caption[ ]{Cumulative mass fraction as a function of look-back time of the solutions obtained for \objectname[]{LGS-3} with different methods, indicated in the label. All them use the DOLPHOT photometry. IAC-B corresponds to the solution adopted and discussed in this paper, obtained with IAC-star/MinnIAC/IAC-pop using the BaSTI stellar evolution library. IAC-G has been obtained with IAC-star/MinnIAC/IAC-pop but using the \citet{gir_etal2000} stellar evolution library. Cole and Match methods make use of the \citet{gir_etal2000} stellar evolution library.
\label{f11}}
\end{figure}

Figure \ref{f11} shows the cumulative mass 
fraction of \objectname[]{LGS-3} as obtained with the three tested methods. 
The agreement between the three methods resulting in the same cumulative mass fraction 
(0.8) at ~8 Gyr is very good. This strongly supports the conclusion the majority the star 
formation in \objectname[]{LGS-3} occurred during this initial episode of star formation. 
Also, all three methods find star formation continuing at much lower rates, until the present. 
The similar behavior of the three solutions allows us to be confident that our scientific 
conclusions are independent of the SFH solution method. In particular, the fact that the 
galaxy has formed most of the stars after the reionization epoch is also confirmed and even reinforced. 

Figure \ref{f11} also shows  
effect of using different stellar models by comparing
the IAC-star/MinnIAC/IAC-pop results using two different stellar libraries.
Here we see that the SFH derived using the BaSTI stellar evolution library has
a slightly faster rise compared to the SFH derived using the \citet{gir_etal2000} 
library.  However, otherwise, the two are in very good agreement, and the main 
features of the SFH of \objectname[]{LGS-3} are found in both derived models at roughly
equivalent levels.

\notetoeditor{Please, put Fig. \ref{f11} here.}

An analysis of the actual resolution provided by the Match and Cole methods, 
of the kind of that done for IAC-star/MinnIAC/IAC-pop in \S\ref{secconstraints} 
would be necessary before going deeper in this comparison, which is beyond our 
objectives here. 
In the following we will use only the results obtained with 
IAC-star/MinnIAC/IAC-pop and the BaSTI stellar evolution library. 

\section{The SFH of LGS-3 compared to other LCID galaxies}\label{seclcid}

In this section, we will compare the SFH of \objectname[]{LGS-3} with that of Phoenix 
\citep[the other dSph/dIrr transition galaxy of the LCID sample,][]{hid_etal2009}
and with those 
of Cetus and Tucana \citep[the two dSphs of the sample,][]{mon_etal2010b,
mon_etal2010c}. 

\begin{figure}
\centering
\includegraphics[width=14cm,angle=0]{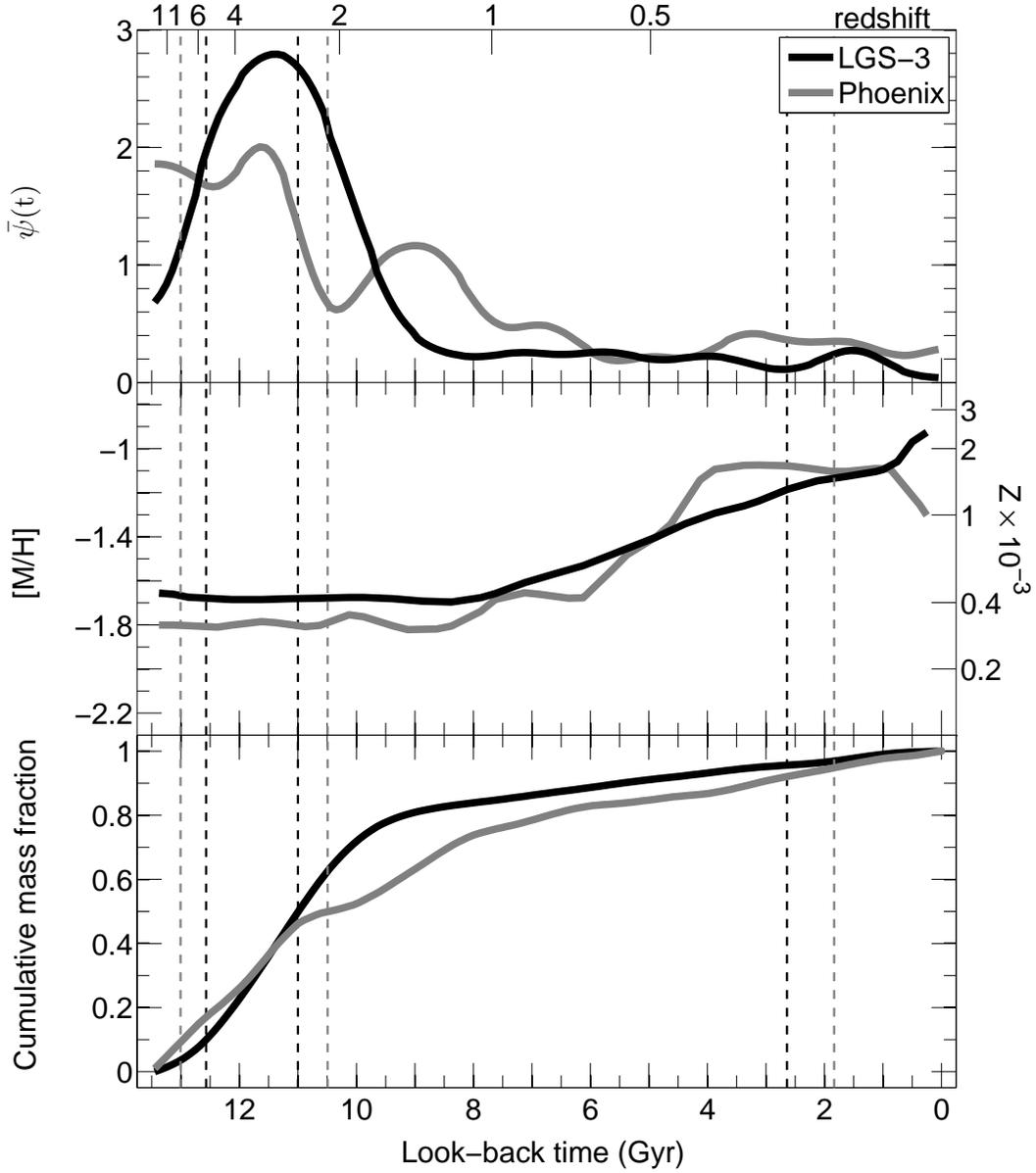}
\protect\caption[ ]{Comparison between the SFHs and the AMRs of LGS-3 
(this paper) and Phoenix \citep{hid_etal2009}. A redshift scale is given 
on the top axis, computed assuming $H_0=70.5\rm~km~s^{-1}~Mpc^{-1}$, $\Omega_m=0.273$, and a 
flat Universe with $\Omega_\Lambda = 1 - \Omega_m$. Vertical dotted-dashed lines 
indicate the times corresponding to the 10$^{\rm th}$, 50$^{\rm th}$ and 95$^{\rm th}$ 
percentiles of $\psi(t)$ for each galaxy, i.e., the times for which the cumulative 
fraction of mass converted into stars was 0.1, 0.5 and 0.95 of the current one. 
\label{f12}}
\end{figure}

Figure \ref{f12} shows the SFH and the AMR of \objectname[]{LGS-3} 
compared to those of Phoenix \citep{hid_etal2009}. 
The correspondence between the main properties of both galaxies is remarkable. 
Both galaxies show a dominant period of initial star formation at early times,
with a result that both galaxies had formed 50\% of their total stellar mass 
about 10.5--11.0 Gyr ago, or before a redshift of 2. 
Both galaxies had also formed more than 95\% of their 
total stellar mass by 2 Gyr ago. 
The AMRs of Phoenix and \objectname[]{LGS-3} show similar behaviors from the earliest times,
with $[M/H]\sim-1.7$, until reaching 95\% of the cumulative mass fraction, 
with $[M/H]\sim-1.2$.  
Note that, in both cases, the slope of the AMR increases about $\sim 8-9$ Gyr ago, 
after the initial episode of star formation. 

\notetoeditor{Please, put Fig. \ref{f12} here.}

The SFHs of both transition type galaxies are similar to those of the two dSphs in 
the sense that an old, main episode exists and that a sharp drop off in the star 
formation rate occurred $\sim 9$ Gyr ago. After this, the transition galaxies, 
Phoenix and \objectname[]{LGS-3}, were able to form stars at low, steadily decreasing rates 
while the dSphs completely stopped their star formation. 
\citet{mon_etal2010c} showed that the first, main episode of star formation 
occurred earlier in Tucana than in Cetus.  The main episode of star formation 
\objectname[]{LGS-3} has an age comparable to that of Cetus and is also clearly younger 
than that of Tucana. 

There are significant 
differences between the AMRs of the transition and the dSph galaxies 
even for $t>9$ Gyr. The metal enrichment from the earliest times  
up to $\sim 9$ Gyr ago is less than 0.1 dex in \objectname[]{LGS-3} and Phoenix, 
while it is more than 0.4 dex in the same period in the dSph galaxies 
\citep{mon_etal2010b,mon_etal2010c}. In the case of the transition galaxies, 
more than 80\% of the stars were formed with roughly the same metallicity. 
This result is supported by the narrow RGB of \objectname[]{LGS-3} as described in 
\S\ref{secred}, indicative of a low metallicity spread for the 
intermediate-to-old age stars. 

The differences between the AMRs of the transition and dSph galaxies is 
intriguing.  Galaxy mass is the fundamental parameter which is thought to 
determine a galaxy's ability to retain gas.  This might naively imply
that the dSphs were initially more massive systems than the transition galaxies.
However, today, Cetus, Tucana, \objectname[]{LGS-3}, and Phoenix all have comparable
luminosities.  Tidal stripping is thought to be very important in 
the evolution of dSph galaxies \citep[see][and references therein]{pen08}.
The comparable present day luminosities, combined with the difference
in AMRs, could point to significantly larger halo masses for the isolated
dSphs which have been lost to tides, or, in other words, a much 
smaller impact of tides on the evolution of the transition galaxies.

\section{Cosmological considerations and the early evolution of LGS-3}\label{seccosmo}

\subsection{Background}

Dwarf galaxies are the focus of a major cosmological problem affecting 
the CDM scenario; the number of dark matter sub-halos 
around Milky Way type galaxies predicted by CDM simulations is an order 
of magnitude larger than observed \citep{kwg93, kly_etal1999, moo_etal1999}. 
Several explanations have been proposed to solve this problem but most of them 
use two main processes that can dramatically affect the formation and evolution 
of dwarf-sized halos: heating from the ultraviolet (UV) radiation arising from 
cosmic reionization and internal supernovae (SNe) feedback. Both are, in principle, 
capable of stopping the star formation in a dwarf halo and even to fully remove 
its gas. The detailed information about the SFH at early and intermediate ages, 
such as we have obtained here, may provide fundamental insight into the problem 
and help to discriminate between the different scenarios proposed so far. 
Detailed discussions about this complex problem can be found in \citet{mac_fer1999}, 
\citet*{bul_etal2001}, \citet{sto_etal2002}, \citet*{kra_etal2004}, 
\citet{ric_gne2005}, \citet{str_etal2008}, and \citet{saw_etal2010}, 
among others. 
Here we will give a short summary just to place our forthcoming discussion in context.

Feedback from SNe can suppress star formation or even blow out 
the gas completely from the smallest systems. \citet{mac_fer1999} presented 
detailed models showing that the mass ejection efficiency is low in galaxies 
with baryonic mass above $\sim 10^7$ M$_\odot$ and that only galaxies with 
baryonic mass below $\sim 10^6$ M$_\odot$ could have their gas completely 
blown away almost independently of the SNe mechanical luminosity. It is 
worth mentioning that, according to the assumptions of \citet{mac_fer1999}, 
a baryonic mass of $\sim 10^6$ M$_\odot$ would correspond to a total mass 
of $6.8\times 10^7$ M$_\odot$, while a baryonic mass of $\sim 10^7$ M$_\odot$ 
would correspond to $3.5\times 10^8$ M$_\odot$. 

There is a consensus that UV background 
heating establishes a characteristic time-dependent minimum mass 
(a {\it filtering} mass) for halos that can accrete gas \citep[e.g.,][]{gnedin00,kra_etal2004}. 
The UV background will heat the gas in low mass halos, preventing star 
formation, and eventually photo-evaporate them. 
As a result, halos not massive enough to have accreted gas before 
the reionization epoch would fail to do so later, unless they accrete 
mass faster than the rate at which the filtering mass increases. 
The red-shift of the 
reionization epoch has been estimated from polarization observations of the 
CMB to be $z=10.9\pm 1.4$ \citep[WMAP 5-year results,][]{kom_etal2009}. 
However, the presence of the Gunn-Peterson trough in quasars at $z\sim 6$ 
shows that the universe was not yet fully re-ionized at earlier epochs 
\citep{loe_bar2001, bec_etal2001}. 

According to models \citep[see e.g.,][]{mac_fer1999}, the minimum 
circular velocity for a dwarf halo to accrete and cool gas in order to produce 
star formation is $v_c\sim30$ km\,s$^{-1}$, which corresponds to a 
total mass of $\sim10^9$ M$_\odot$. 
However, most dwarf galaxies in the Local Group, including all the 
Milky Way satellites and five out of the six of the LCID sample, show 
circular velocities well below that value. 
Nevertheless, there are currently several proposed ways to overcome 
this apparent contradiction. 
\citet{bul_etal2001} has proposed that dSph 
and the rest of the smallest halos would have been formed during 
the pre-reionization era. 
The processes of photoionization feedback resulted in the suppression 
of the star formation below an observable level in 90\% of them. 
This was supported by \citet{ric_gne2005}, who found that 
if positive feedback is considered, a fraction of low mass halos 
(total mass below $10^8$ M$_\odot$) would have been able to form 
stars before the reionization era. 
The star formation in these 
galaxies would have been halted by internal mechanisms, like 
photo-dissociation of H$_2$ or SNe feedback in advance of the 
reionization era.
Another possibility is that masses of dark matter halos of the dSphs 
galaxies would be much larger than those measured at the optical limit 
of the galaxies and larger than the limit required by the reionization 
scenario \citep{sto_etal2002}. 
In these scenarios, dark matter halos of lower masses would have 
failed to form stars and remain dark. 
Alternatively, dark matter halos could have been much larger in the 
reionization epoch, with total masses above the mass required for 
star formation, but they lost a large fraction of their mass afterward 
due to tidal harassment \citep{kra_etal2004}. It is also possible that
a self-shielding mechanism would be present, protecting the gas in the 
inner regions or in regions denser than some limit from being heated by the 
UV background \citep{sus_ume2004}.
Finally, it has been suggested that inhomogeneous reionization can lead to 
large variations in the masses of halos that survive and produce stars
\citep{busha10}.

In summary, it is useful to note that the first scenario  
\citep{bul_etal2001,ric_gne2005} and also the last ones \citep{sus_ume2004, 
busha10}, 
allow star formation in low mass halos (total mass below $10^8-10^9$ M$_\odot$), 
while the second and third are compatible with star formation in more massive 
halos only (total mass above $10^9$ M$_\odot$). Additionally, the first scenario 
would imply that star formation would have been produced before 
reionization and would have been halted afterward in the smallest galaxies. 

Recently, \citet{saw_etal2010} have presented a set of high-resolution  
hydrodynamical simulations of the formation and evolution of isolated dwarf galaxies 
including the most relevant physical effects, namely metal-dependent cooling, 
star formation, feedback from Type II and Ia SNe, and UV background radiation. 
Their results are very useful for a direct comparison with observations. 
They study halos with present day total masses between $2.3\times 10^8$ M$_\odot$ 
and $1.1\times 10^9$ M$_\odot$ and reach an interesting set of conclusions 
that include many of the aforementioned ones. First, halos that are not 
massive enough to accrete sufficient gas to form stars before $z=6$ lose 
their gas subsequently due to UV background heating. 
In this sense, reionization sets a lower mass limit of the halo for the star formation. 
Second, in halos that form stars, feedback is the main process driving 
the evolution of the galaxy and regulating its star formation. 
It alone can blow out all the gas of a dwarf system before $z=0$, 
while UV background alone has almost no effect. 
However, if feedback has previously made the gas diffuse and reduced 
its radiative cooling efficiency, then the UV background has a strong effect, 
producing a sharp cut off in the star formation at the epoch of reionization. 
Finally, self-shielding is effective only in the inner regions of the halos 
at the more massive end.

All the aforementioned mechanisms are founded on solid and extensive 
theoretical modeling. However to decide to what extent one or another 
should be preferred, detailed observational evidence is necessary. 
One such observable is whether the star formation in the smallest galaxies 
was halted at reionization. 
To answer this question, a precision of $\sim 1$ Gyr is required in 
the determination of the SFH at $12.5-13$ Gyr ago \citep{ric_gne2005}, 
which is indeed what we have reached in our work. 
In the following we will discuss the different possibilities in light 
of our results for \objectname[]{LGS-3} paying attention first to SNe feedback effects 
and second to UV background effects. 

\subsection{SNe feedback in LGS-3}\label{secfeedback}

The SFH obtained for \objectname[]{LGS-3} allows us to make a direct estimate of the importance 
of feedback in \objectname[]{LGS-3}. To do so, the mechanical luminosity of the SNe produced 
in the main star formation episode and the mass of the galaxy can be compared 
with the results of \citet{mac_fer1999}. They used models to calculate 
the conditions for blow-away, partial mass loss, and no mass loss of 
dwarf galaxies as a function of the baryonic mass of the galaxy and the 
mechanical luminosity, $L_w$, of the SNe produced during a central,
instantaneous star formation episode. 

The kinetic energy released to the interstellar medium by the SNe produced 
in the old, main star formation episode in \objectname[]{LGS-3}, can be computed as follows. 
For a continuous star formation rate of 1 M$_\odot$ yr$^{-1}$,
the STARBURST99 models \citep{lei_etal1999} predict a
mechanical luminosity of $\log L_w=41.9$.
This is the equilibrium value reached after a few tens of million years 
from the first star formation. The star formation rate is for a mass range 
1 to 100 M$_\odot$ and is computed using a classical IMF with exponent $-2.35$. 
For direct comparison, we convert this rate to the same 
stellar mass interval and IMF that we have used and obtain 1.7 M$_\odot$ yr$^{-1}$.
Scaling these values to the SFR obtained for \objectname[]{LGS-3}, the total mass converted 
into stars in the initial episode is $1.7\times 10^6$ M$_\odot$. 
Assuming a Gaussian profile of $\sigma=0.6$ Gyr, or a duration (FWHM) of 1.4 Gyr,
results in an average SFR of $1.3\times 10^{-3}$ M$_\odot$ yr$^{-1}$. 
Scaling STARBURST99 models to this average SFR results in a mechanical
luminosity of $L_w=5.3\times 10^{38}$ erg s$^{-1}$.

Alternatively, we can compute the mechanical luminosity from core collapse SNe only
(note that STARBURST99 includes all sources of mechanical luminosity). From  
\S\ref{secsfhmethod}, the total number of SNe formed during the old, main episode 
was $2.9\times 10^4$ or $1.6\times 10^4$ for minimum core collapse progenitor 
masses of 6.5 M$_\odot$ or 10 M$_\odot$, respectively
\citep{sal_cas2005}. 
Assuming an energy release per SN of $10^{51}$ erg \citep{lei_etal1999}, 
we obtain a total mechanical luminosity released during the old, main episode of 
$L_w=6.8\times 10^{38}$ erg s$^{-1}$ or $L_w=3.8\times 10^{38}$ erg s$^{-1}$, 
respectively. Since these values are compatible with the STARBURST99 
estimate, we will adopt $L_w=5.3\times 10^{38}$ erg s$^{-1}$.
Combining this with a baryonic mass of $1.5\times 10^6$ M$_\odot$ 
(\S\ref{secsfhfeatures}), the model results of \citet{mac_fer1999}
shown in their Figure 1, place \objectname[]{LGS-3} in the regime of mass
loss, but fairly close to the blow-away regime.

However, equation 1 from \citet{mac_fer1999} gives a relation between the baryonic 
and dark matter mass of dwarfs which is an extrapolation to low mass 
galaxies of the relation derived by \citet*{per_etal1996} for more massive 
galaxies (total masses above $10^{10}$ M$_\odot$). 
From this relation, the total mass for \objectname[]{LGS-3} is $9\times 10^7$ M$_\odot$. 
Alternately, \cite{saw_etal2010} give the final ($z=0$) stellar, gaseous, and 
total masses for their model runs. The relations between these parameters
depend on the model ingredients, but for a baryonic mass like that of \objectname[]{LGS-3}, 
the total mass is in the range from $2\times 10^8$ M$_\odot$ to 
$4\times 10^8$ M$_\odot$.  Using this estimate instead of the 
\citet{per_etal1996} extrapolation places \objectname[]{LGS-3} well within the regime 
of mass-loss due to feedback and far from the blow-away regime.

\subsection{The effects of reionization}\label{secreionization}

A flat Einstein-de Sitter universe with $H_0=70.5\ \rm km~s^{-1}~Mpc^{-1}$, 
$\Omega_m=0.273$ \citep[as derived from the WMAP 5-year data;][]{kom_etal2009}, 
has an age $\sim 13.7$ Gyr.  The WMAP 5-year data indicate the 
epoch of reionization occurred at $z\sim 10.9$ 
corresponding to a look-back time of $\sim 13.3$ Gyr.
Evidence from quasar spectra indicate the epoch of reionization is
over at a redshift of $z\sim 6$ or a look-back time of $\sim 12.7$ Gyr
\citep{bec_etal2001}.  We will focus on the latter since it is more 
conservative from the point of view of the forthcoming discussion. 

Did \objectname[]{LGS-3} form most of its stars before or after the epoch of reionization?
Using the cumulative mass fraction from Figure 
\ref{f7}, 
we find that more than 80\% of the stellar mass of \objectname[]{LGS-3} has been 
formed later than $\sim 12.5$ Gyr ago ($z\sim 5$).  Thus, the observations
indicate that the majority of the stars in \objectname[]{LGS-3} were formed after the 
epoch of reionization.  

\begin{figure}
\centering
\includegraphics[width=14cm,angle=0]{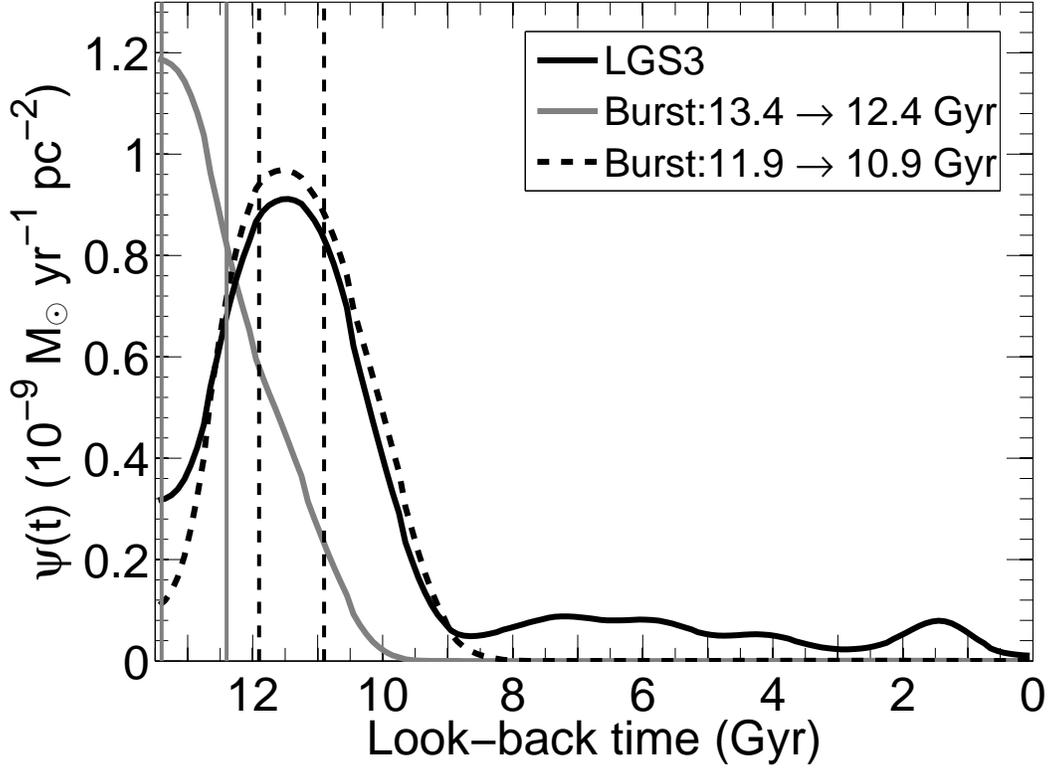}
\protect\caption[ ]{The SFH obtained for two mock galaxies, compared with the actual SFH of \objectname[]{LGS-3}. One of the mock galaxies simulates a system in which most or all the star formation occurred before reionization (star formation between 12.4 and 13.4 Gyr ago; magenta, dotted line). The other one simulates a system in which all the star formation occurred after reionization (star formation between 10.9 and 11.9 Gyr ago; black, dotted line). Magenta and black solid lines show the solutions obtained for them while the green solid line correspond to \objectname[]{LGS-3}. This figure provides strong evidence that most, if not all, the star formation in \objectname[]{LGS-3} occurred after reionization. 
\label{f13}}
\end{figure}

We can use synthetic data to better assess this constraint.  
In Figure \ref{f13}, we show the results of
two synthetic SFHs, one in which all stars are formed between 13.4 and 
12.4 Gyr ago (before the epoch of reionization) and another where all of
the stars are formed between 11.9 and 10.9 Gyr ago.
Figure \ref{f13} shows that for the model where
all of the stars formed before the epoch of reionization is clearly
inconsistent with the SFH of \objectname[]{LGS-3}.  On the other hand, the model with
all stars forming after the epoch of reionization is a good match to 
our SFH for \objectname[]{LGS-3}, indicating that it is highly unlikely that \objectname[]{LGS-3}
formed a large fraction of its stars before reionization had concluded.

\notetoeditor{Please, put Fig. \ref{f13} here.}

However, the absolute ages in our comparisons are intrinsically dependent 
on the stellar evolution models which have intrinsic uncertainties
\citep[e.g.,][]{chaboyer95, chaboyer98}.  Modern stellar evolution
models have addressed many of the shortcomings in earlier generations of 
models \citep[e.g., see discussion in][]{dotter07}, and the results of
the relative ages derived from main sequence stars are in excellent 
agreement (better than 0.5 Gyr) as demonstrated by \citet{mar_etal2009}.  
The accuracy of the absolute ages is also increasing.  For example, 
the average age of the oldest globular clusters obtained 
from the age analysis by \citet{mar_etal2009}, 
using the BaSTI library, is 12.3 Gyr with an age dispersion 
of $\pm 0.4$ Gyr (Mar\'\i n-Franch, private communication).
In this regard, the complete inconsistency of our SFH for \objectname[]{LGS-3} 
with the model in which all stars are formed before reionization
is persuasive.

Nonetheless, as argued in \citet{mon_etal2010c}, the strongest constraints
come from relative ages.  Because all of the LCID galaxies were observed and
analyzed in an identical manner we can make the following argument. Since 
we find an age difference of $>1.5$ Gyr between the peak of star formation 
in \objectname[]{LGS-3} and that of Tucana, and since that difference is larger than the
difference between the age of the universe and the epoch of reionization,
it is highly unlikely that a significant fraction of the stars in
\objectname[]{LGS-3} have been formed before the epoch of reionization.
An identical conclusion was found for the Cetus dSph \citep{mon_etal2010b}.
In summary, Figure \ref{f5f6} provides strong 
evidence that most, if not all, of the star formation in \objectname[]{LGS-3} occurred 
after the end of reionization.

Although \objectname[]{LGS-3} is a low mass galaxy, do theoretical models expect it 
to form its star before reionization?
Before putting the former result in the context of current theoretical 
model predictions, some estimates of integrated physical quantities 
of \objectname[]{LGS-3} are necessary. The velocity dispersion of the gas in \objectname[]{LGS-3} 
is 8 km\,s$^{-1}$ at a radius of 470 pc \citep{you_lo1997} which corresponds 
to a total mass within that radius of $2.1\times 10^7$ M$_\odot$. 
According to \citet{sto_etal2002} this value should {\it not} be considered 
indicative of the total mass of the galaxy, but only of the virial mass 
at the measured radius. 
Following the results of \cite{saw_etal2010}, 
the total mass of \objectname[]{LGS-3} would be expected to be in the interval 
$\sim2\times 10^8$ M$_\odot$ to $\sim4\times 10^8$ M$_\odot$. 
Equation 16 of \citet{mac_fer1999} gives a relation between 
circular velocity and total mass in a density distribution 
given by an modified isothermal sphere. 
Using $h=0.7$, it can be written as $v_c(r)=10.5\ M_{h,7}^{1/3}$, 
where $M_{h,7}$ is the total mass in units of $10^7$ M$_\odot$ 
and $v_c$ is given in km\ s$^{-1}$. 
Thus, for a total mass in the range of $2\times 10^8$ M$_\odot$ to 
$\sim4\times 10^8$ M$_\odot$, the resulting circular velocity for \objectname[]{LGS-3} 
is between 28 km\ s$^{-1}$ and 36 km\ s$^{-1}$. 
These values are close to or somewhat above the limit of 
30 km\ s$^{-1}$ below which the UV background is expected to prevent any 
subsequent star formation.

In summary, it is possible that \objectname[]{LGS-3} was massive enough that its star formation  
was not halted by UV background heating and it was also able to preserve at 
least part of its gas against feedback from SNe. 
However, models by \citet{saw_etal2010} indicate that, for its estimated
mass, the {\it combination}
of the the UV background together with SN feedback should have halted 
the star formation at $z\sim 6$ (or $\sim 12.7$ Gyr ago). 
Since this is evidently not the case, a mechanism such as self-shielding 
seems necessary to allow the galaxy continue forming stars, at least 
in its inner regions, as suggested by \citet{saw_etal2010}. 
If the total mass of \objectname[]{LGS-3} would be of the order of $10^8$ M$_\odot$ or 
lower \citep[as obtained from the relation used by][]{mac_fer1999}, 
the circular velocity would drop to values of the order of 20 km\ s$^{-1}$. 
For this total mass estimate, our observation that most of the star 
formation in \objectname[]{LGS-3} has taken place after $z\sim 6$ would be difficult 
to reconcile with models.  However, one possibility is that  
a mechanism like mass loss due to tidal harassment \citep[as proposed 
by][]{kra_etal2004} would be at play. 
On the other hand, the \citet{mac_fer1999} 
relation requires a large extrapolation for the mass of \objectname[]{LGS-3}, and
this may simply imply that the \citet{saw_etal2010} mass 
estimates are to be preferred.

\section{Summary and Conclusions}\label{seccon}

We have presented an analysis of the SFH of the transition-type dwarf 
galaxy \objectname[]{LGS-3}, based on deep HST photometry obtained with the ACS. 
The assumed distance of the galaxy is 652 kpc \citep{ber2009}. 
A summary of the results follows.

\begin{itemize}

\item A deep CMD reaching the oldest main-sequence turn-offs with 
completeness of $\sim 0.90$ has been obtained. Exhaustive crowding 
tests have been performed in order to properly characterize the 
observational effects. Careful simulations of the observational effects
have been conducted, taking into account the variable amount of 
crowding within the galaxy. 

\item The SFH for the entire lifetime of the galaxy has been obtained.
Tests of self-consistency, uniqueness, and stability of the solution 
have been performed together with tests exploring the dependency of 
the solution on the photometric reduction package, 
stellar evolution library, and the SFH derivation code.

\item The solution shows that the SFH of \objectname[]{LGS-3} is dominated by an
old, main episode with maximum occurring $\sim 11.7$ Gyr ago and a duration 
estimated in 1.4 Gyr (FWHM).  Subsequently, \objectname[]{LGS-3} has continued forming 
stars until the present at a much lower rate.

\item The total mass of stars produced is $2.0\times 10^6$ M$_\odot$. 
Of that, $1.7\times 10^6$ M$_\odot$ corresponds to the old, main star 
formation episode. The current mass in stars and stellar remnants is 
$1.1\times 10^6$ M$_\odot$. Using a gas mass of 
$3.8\times 10^5$ M$_\odot$ \citep{you_lo1997} the total baryonic mass 
of the galaxy is $1.5\times 10^6$ M$_\odot$.

\item There is little evidence of chemical enrichment during the first 
main episode of star formation lasting until $\sim 9$ Gyr ago. 
After that, the metallicity increased more steeply to a present day value 
of $Z \sim 0.0025$. 
The epoch in which the enrichment rate increased is coincident with 
the epoch at which the star formation drops to a low value. 
This suggests a scenario in which \objectname[]{LGS-3} formed stars mainly from 
fresh gas during the first part of its life. 
After $9$ Gyr ago much of the gas supply was exhausted and stars formed 
from a larger fraction of recycled gas. This resulted in both a lower star 
formation rate and an increase of the chemical enrichment rate.

\item The difference between the lack of early chemical enrichment in 
the isolated transition galaxies compared to the significant early 
chemical enrichment in the isolated dSph galaxies of comparable
luminosities may be indicative that the dSph galaxies experienced
significantly more tidal stripping.

\item Most or all the star formation was produced in \objectname[]{LGS-3} after the 
reionization epoch, assumed to occur $\sim 12.7$ Gyr ago.

\item The mass of \objectname[]{LGS-3} and the mechanical 
luminosity from SNe associated with the old, main episode 
($L_w=5.3 \times 10^{38}$ erg s$^{-1}$) indicate that  
mass-loss by galactic winds is important, but that 
complete blow-away was not likely according to  
the models of SN feedback by \citet{mac_fer1999}. In other words, 
\objectname[]{LGS-3} seems to be massive enough to 
conserve at least a fraction of its gas against SNe feedback.

\item According to models by \citet{saw_etal2010}, the total mass of 
\objectname[]{LGS-3} is about $2\times 10^8$ M$_\odot$ to $4\times 10^8$ M$_\odot$, 
which corresponds to a circular velocity of 28 km\ s$^{-1}$ to 36 km\ s$^{-1}$ 
\citep{mac_fer1999}. These values are close to or somewhat above the 
limit of 30 km\ s$^{-1}$ below which the UV background radiation would 
prevent any subsequent star formation.

\item According to \citet{saw_etal2010}, the combined effects of 
feedback and UV background radiation could have stopped the star 
formation in a galaxy like \objectname[]{LGS-3} after reionization, unless some other 
mechanism like self-shielding is present. Indeed, comparison of the 
SFH of \objectname[]{LGS-3} with model results by \citet{saw_etal2010} indicate that 
self-shielding should be at play to prevent complete cessation of star 
formation in \objectname[]{LGS-3} at the epoch of reionization.

\end{itemize}

The results presented in this paper allow us to sketch the following scenario for the 
early evolution of \objectname[]{LGS-3}.  The peak of star formation 
occurred $\sim 11.7$ Gyr ago, and the star formation rate dropped 
dramatically $\sim 9$ Gyr ago or earlier although a low star formation 
activity is observed until present. 
Most or all the star formation occurred in 
the galaxy after the end of reionization (although some star formation 
before that time cannot be ruled out). Feedback from SNe would have 
created outflows, but would have been insufficient to completely blow-away the gas. 
The drop in star formation at $\sim 9$ Gyr ago may have been due to the 
combined results of feedback, UV background heating, and gas exhaustion 
due to star formation. 
According to models presented by \citet{saw_etal2010}, at the epoch of 
reionization, (in the absence of self-shielding),
the combination of heating by cosmic UV-background and SNe feedback should have 
completely stopped the star formation in the galaxy. 
The presence of the star formation activity extending to
the present leads us to infer that a self-shielding mechanism has been 
important in the early evolution of \objectname[]{LGS-3}. 

\acknowledgments

We are very grateful to Dr. C. P\'erez Gonz\'alez and C. Mu\~noz-Tu\~n\'on 
for fruitful comments. 
The computer network at IAC operated under the Condor software license has been used. 
Authors SH, AA, CG and MM are funded by the IAC (grant P3/94) and by the Science 
and Technology Ministry of the Kingdom of Spain (grant AYA2007-3E3507). 
SC is funded by the Science and Technology Ministry of the Kingdom of Spain 
(grant AYA2007-3E3507). 
Support for this work was provided by NASA through grant GO-10515
from the Space Telescope Science Institute, which is operated by
AURA, Inc., under NASA contract NAS5-26555.
This research has made use of NASA's Astrophysics Data System
Bibliographic Services and the NASA/IPAC Extragalactic Database
(NED), which is operated by the Jet Propulsion Laboratory, California
Institute of Technology, under contract with the National Aeronautics
and Space Administration.

\appendix

\section{Confidence of the solution}\label{secimf}

As we have mentioned in \S\ref{secsfhmethod}, solutions which are unique, stable, 
self-consistent, and as independent as possible of the choice of parameters and model functions are desirable. We have done many tests to evaluate these properties 
in the SFHs derived for LCID galaxies.  Here we briefly discuss these aspects
of the calculated SFHs.

\subsection{Uniqueness}\label{uniqueness}

This condition is fulfilled if combinations of simple populations which are within the error bars of the adopted solution produce CMDs which are indistinguishable 
from the best fit CMD. It is also required that combinations of simple populations significantly different from the adopted solution  produce calculated CMDs significantly different from the solution one. Simulations done during the tests for the IAC-star/MinnIAC/IAC-pop method \cite[][and this work]{apa_hid2009, hid_etal2009}, consisting in obtaining the SFH of mock stellar population of known SFH, show that the aforementioned conditions are fulfilled in general.

\subsection{Self-consistency}\label{cons}

By self-consistency we refer to the capability of the method to reproduce 
synthetic input. 
When a mock population is computed according to a given SFH and analyzed 
in an identical manner as real data, the resulting SFH should be equal to the input one, within errors and resolution limits. 
Such tests have been executed in \citet{apa_hid2009} and repeated here 
using the best fit SFH given in Fig. \ref{f5f6} to compute the mock population 
to be analyzed. 
For simplicity, we will show here only the tests for DOLPHOT photometry but 
the result is equivalent for DAOPHOT photometry. 
See also tests presented in \S\ref{secconstraints}.

\begin{figure}
\centering
\includegraphics[width=14cm,angle=0]{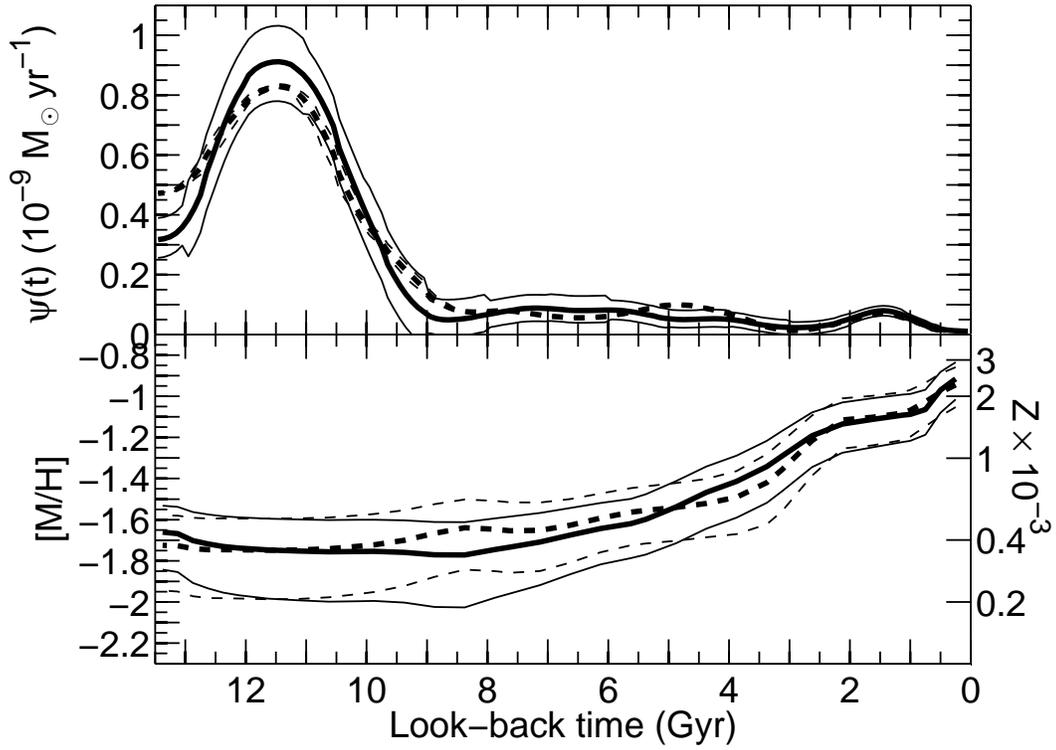}
\protect\caption[ ]{Consistency test of the SFH of \objectname[]{LGS-3}. Thick solid line gives the solution SFH and metallicity of the galaxy using DOLPHOT photometry. Thick dashed line shows the recovered SFH when the solution SFH is used as mock SFH input. Thin lines show error intervals in top panel and real dispersion in bottom panel. 
\label{f14}}
\end{figure}

To minimize the dependency of the test results on the associated statistical 
fluctuations inherent to the method (present in the observational 
uncertainties simulation and in the synthetic CMD modeling), we have 
obtained the SFHs for 10 CMDs calculated from the SFH of Figure \ref{f5f6} 
changing only the random number generation seed. An example is shown in 
Figure \ref{f8}. Figure \ref{f14} shows the average of all of the new SFHs 
and the AMRs obtained. The input values, i.e., the SFH and AMR of 
\objectname[]{LGS-3}, are also shown for comparison. The main conclusion 
of the self-consistency test is that the SFH recovering process 
introduces no significant bias in the location of the main peak nor in the AMR, 
and that the SFH features are properly recovered, although it introduces 
some degree of degradation in time resolution. This has to be taken into 
account and corrected for when quantities like the maximum intensity of a burst or its duration are relevant (see \S\ref{secconstraints}).

\notetoeditor{Please, put Fig. \ref{f14} here.}

\subsection{Stability}\label{stab}

Stability is related to uniqueness, but the question we want to address 
here is what is the maximum excursion of  the model functions and 
external parameters producing a SFH within the error bars of the solution. 
We have tested the effects of changing the fraction of binary stars, 
the IMF, and the photometric zero points. The latter are discussed 
in \S\ref{secsfhresult}. The dependency of the solution on the fraction 
of binary stars and the IMF has been analyzed in detail in 
\citet{mon_etal2010b} and E.\ Skillman et al.\ (2011, in prep.) respectively. 
In short, any fraction of binary stars larger than 0.4 gives a SFH 
within the error bars. In the case of the IMF, for the interval 
of stellar masses $0.5<m\leq 100~ M_\sun$, the range of values 
producing solutions within the errors is $2.2\leq x\leq 2.4$, 
where $x$ is the mass exponent index in $\phi(m) = A m^{-x}$. 
The solution is independent of $x$ for the interval $0.1\leq m\leq 0.5~ M_\sun$, 
although the total stellar mass is not. As an indication, the total mass 
converted into stars in \objectname[]{LGS-3} would be increased by a 
factor $\sim 1.5$ if the IMF exponent changes from $x=-1.3$ to $x=-2.3$ 
for the interval $0.1\leq m\leq 0.5~ M_\sun$. The same factor of 
$\sim 1.5$ exists between the corresponding current masses in existing 
stars and stellar remnants. As a summary, Table \ref{parrange} shows the 
range of variation for the binary fraction, IMF, and photometric offsets 
which give solutions within the error bars for the adopted 
SFH of \objectname[]{LGS-3}. 

\notetoeditor{Please, put Table \ref{parrange} here.}

\subsection{Statistical significance}\label{sign}

\begin{figure}
\centering
\includegraphics[width=14cm,angle=0]{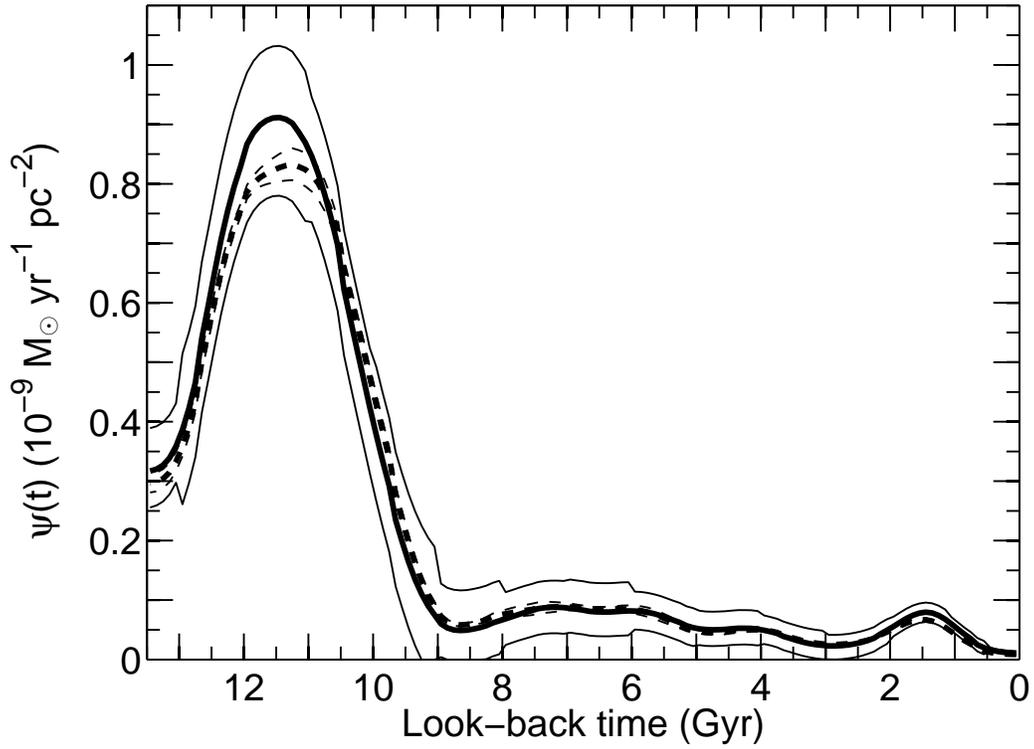}
\protect\caption[ ]{Statistical significance of the SFH of \objectname[]{LGS-3}. Solid lines shows the SFH of \objectname[]{LGS-3} and its error interval. Thick dashed line shows the average of the SFHs of five subsets of stars of \objectname[]{LGS-3} each one containing a 50\% of stars randomly selected. The results of the
tests were normalized by scaling by a factor of 2. 
Thin dashed lines show the rms deviations of the five solutions. 
\label{f15}}
\end{figure}

Since the number of real stars is limited, the observational sampling 
could be insufficient and could introduce undesirable biases in the 
solution. To check this point we have obtained the SFH of five different 
random subsets of the observed stars. Each one contains 50\% of the total, 
so that each star may have been included in several subsamples. 
Figure \ref{f15} shows the SFH of \objectname[]{LGS-3} and the average 
of the SFHs of the five subsamples. For the comparison to be meaningful, 
the test results have been normalized by a factor.  Error bars show the 
dispersion of the five subsample SFHs. The agreement is very good and 
no bias is apparent, which indicates that our solution does not suffer 
from under-sampling effects. 

\notetoeditor{Please, put Fig. \ref{f15} here.}

\clearpage

\clearpage

\clearpage

\clearpage

\clearpage

\clearpage

\clearpage

\clearpage

\clearpage

\clearpage

\clearpage

\clearpage

\clearpage

\clearpage

\begin{deluxetable}{ccc}
\tabletypesize{\scriptsize}
\tablecaption{Summary of observations\label{logobs}}
\tablewidth{0pt}
\tablehead{
\colhead{Date / UT} &\colhead{Exp. Time (s)} &\colhead{Filter}}
\startdata          
2005-09-12 16:18:15 &2512 &F475W\\
2005-09-12 16:42:00 &2304 &F814W\\
2005-09-14 16:16:02 &2512 &F475W\\
2005-09-14 16:39:47 &2304 &F814W\\
2005-09-15 14:38:40 &2512 &F475W\\
2005-09-15 15:02:25 &2304 &F814W\\
2005-09-16 13:01:46 &2512 &F475W\\ 
2005-09-16 13:25:31 &2304 &F814W\\
2005-09-17 11:24:47 &2512 &F475W\\	 
2005-09-17 11:48:32 &2304 &F814W\\
2005-09-17 14:36:38 &2512 &F475W\\ 
2005-09-17 15:00:23 &2304 &F814W\\
\enddata
\end{deluxetable}
\clearpage

\begin{deluxetable}{lcccc}
\tabletypesize{\scriptsize}
\tablecaption{Functions and parameters tested and adopted for the SFH\label{varpar}}
\tablewidth{0pt}
\tablehead{
\colhead{Function/Parameter} &\colhead{Min. value} &\colhead{Max. value} &\colhead{Step} &\colhead{Adopted}}
\startdata
Stellar evolution library                              &BaSTI   &Girardi &\nodata &BaSTI\\
Binary fraction                                         &0.0    &1.0     &0.2    &0.4\\
IMF ($\rm 0.1\leq m\leq 0.5~ M_\odot$)\tablenotemark{a} &0.0    &4.0     &0.1    &1.3\\
IMF ($\rm 0.5<m\leq 100~ M_\odot$)                      &0.0    &5.0     &0.1    &2.3\\
$\Delta Z_{F475W}$\tablenotemark{b}                     &-0.21  &0.21    &0.105  &0.13/0.08\\
$\Delta Z_{F814W}$\tablenotemark{b}                     &-0.15  &0.15    &0.075  &0.11/0.11\\
CMD gridding\tablenotemark{c,d}   &$0.01\times 0.2$  &$0.01\times 0.2$ &\nodata &\nodata \\
Age sampling ($\rm\leq 1~Gyr$)\tablenotemark{d}  &0.2 Gyr           &0.5 Gyr &\nodata &\nodata \\
Age sampling ($\rm> 1~Gyr$)                      &1.0 Gyr           &1.0 Gyr &\nodata &\nodata \\
Metallicity sampling ($\rm Z\leq 0.001$)\tablenotemark{d} &0.0002  &0.0002  &\nodata &\nodata \\
Metallicity sampling ($\rm Z>0.001$)                      &0.0005  &0.0005  &\nodata &\nodata \\
\enddata

\tablenotetext{a}{Values are the exponent {\it x} in the IMF defined as $\phi(m)=A~m^{-x}$, where {\it m} is the mass and {\it A} is a normalization constant.}
\tablenotetext{b}{Zero point offsets are relative to \citet{sir_etal2005}. The adopted values were selected by interpolation. Results for DOLPHOT/DAOPHOT are given in the last column.}
\tablenotetext{c}{Values are the box size (color$\times$magnitude) for the CMD sampling.}
\tablenotetext{d}{Several bin sets were tested using the same bin size but applying different offsets to the starting point of the bin set.}
\end{deluxetable}
\clearpage

\begin{deluxetable}{cccc}
\tabletypesize{\scriptsize}
\tablecaption{Summary of integrated and mean quantities of the LGS-3 SFH\label{meansolpar}}
\tablewidth{0pt}
\tablehead{
\colhead{Area (kpc$^2$)} &\colhead{Mass formed($10^6$ M$_\odot$)} &\colhead{$<$[M/H]$>$} &\colhead{$<\psi(t)> (10^{-3}$ M$_\odot$ yr$^{-1})$}}
\startdata     0.42  &$1.26\pm 0.04$    &$-1.7\pm 0.1$ &$0.09\pm 0.03$\\
   Total Galaxy           &$1.95\pm 0.04$    &$-1.7\pm 0.1$ &$0.14\pm 0.03$\\
\enddata
\end{deluxetable}
\clearpage

\begin{deluxetable}{lc}
\tabletypesize{\scriptsize}
\tablecaption{Model functions and external parameters best fit range\label{parrange}}
\tablewidth{0pt}
\tablehead{
\colhead{Function/Parameter} &\colhead{Min./Max. value}}
\startdata
Binary fraction                                         &0.4/0.8\\
IMF ($\rm 0.1\leq m\leq 0.5~ M_\odot$)\tablenotemark{a} &-4.0/0.0\\
IMF ($\rm 0.5<m\leq 100~ M_\odot$)                      &-3.0/-2.0\\
$\Delta Z_{F475W}$                                      &-0.04/0.02\\
$\Delta Z_{F814W}$                                       &-0.06/0.04\\
\enddata
\tablenotetext{a}{Values are the exponent {\it x} in the IMF defined as $\phi(m)=A~m^{-x}$, where {\it m} is the mass and {\it A} is a normalization constant.}
\end{deluxetable}
\clearpage

\end{document}